\documentstyle[12pt]{article}
\font\srm=cmr9

\def\be{\begin{equation}}
\def\fe{\end{equation}}

\def\ssum{ {\hbox{$\sum\,$}} }

\def\spose#1{\hbox to 0pt{#1\hss}}\def\lta{\mathrel{\spose{\lower 3pt\hbox
{$\mathchar"218$}}\raise 2.0pt\hbox{$\mathchar"13C$}}}  \def\gta{\mathrel
{\spose{\lower 3pt\hbox{$\mathchar"218$}}\raise 2.0pt\hbox{$\mathchar"13E$}}}

\begin{document}

\title{\bf Covariant analysis of Newtonian multi-fluid
models for neutron stars: III Transvective, viscous, 
and superfluid drag dissipation.}

\author { {\bf Brandon Carter \& Nicolas Chamel }\\
 \hskip 1 cm\\   \\Observatoire de
Paris, 92195 Meudon, France.}

\date{\it October, 2004}

\maketitle

\vskip 1  cm

{\bf Abstract. } As a follow up to articles dealing firstly with a convective
variational formulation in a Milne-Cartan framework for non-dissipative
multi fluid models, and  secondly with various ensuing stress energy
conservation laws and generalised virial theorems, this work continues a
series showing how analytical procedures developed in the context of
General Relativity can be usefully adapted for implementation in a purely
Newtonian framework where they provide physical insights that are not so
easy to obtain by the traditional approach based on a 3+1 space time
decomposition. The present article describes the 4-dimensionally covariant
treatment of various dissipative mechanisms,  including viscosity
in non-superfluid constituents, superfluid vortex drag, ordinary resistivity 
(mutual friction) between relatively moving non-superfluid constituents, and 
the transvective dissipation that occurs when  matter is transformed from 
one constituent to another due to chemical disequilibrium such as may be 
produced by meridional circulation in neutron stars. The corresponding 
non dissipative limit cases of vortex pinning, convection, and chemical 
equilibrium are also considered.

\vfill\eject

\bigskip
{\bf 1. Introduction}
\medskip

This article continues the development~\cite{CCI,CCII} of a coherent
fully covariant approach to the construction and application of Newtonian
fluid models of the more general kind required in the context of neutron
star phenomena in cases for which it is necessary to allow for independent
motion of neutronic and protonic constituents. Generalising the approach
that was originally introduced ~\cite{CK94} for the special case 
of Landau's two-constituent superfluid model, the preceding articles 
dealt with idealised perfectly conservative models, for which a 
strictly variational formulation is available. In a complementary
treatment by Prix~\cite{Prix04}, it has been shown how this covariant 
variational formulation can be translated into terms of the 
familiar kind of 3+1 direct product structure of space with time  
that has traditionally be used in a non-relativistic approach,
and also how it can be extended to allow for dissipative effects --
such as chemical reactions and mutual resistivity between relatively 
moving currents of thermal and other kinds -- while retaining as much as 
possible of the convenient~\cite{PCA,YE04} mathematical machinery provided 
by the variational approach. In the same spirit, but continuing within a 
fully covariant framework, the purpose of the present article is to 
extend our treatment to allow for a wider range of dissipative 
mechanisms of the kind~\cite{LM00} likely to be relevant in neutron stars, 
particularly those due to the presence of superfluid vortices.

In the non-dissipative applications considered in the preceding 
articles~\cite{CCI,CCII} the status of any entropy current 4-vector 
$s^\mu$ that may have been involved was effectively the same as that of 
the other relevant conserved currents with 4-vectors $n_{_{\rm X}}^{\,\mu}
=n_{_{\rm X}}u_{_{\rm X}}^{\,\mu}$ for corresponding  number density
$n_{_{\rm X}}^{\,\mu}$ and unit flow vectors $u_{_{\rm X}}^{\,\mu}$
designated by various values of the chemical index label {\srm X}.
However, in the dissipative applications to be considered here, the
entropy density $s$ and the (no longer conserved) entropy current
\be s^\mu=s u_\emptyset^{\,\mu}\label{1}\fe
will have a privileged role, characterising a corresponding local
{\it thermal rest frame} specified by a unit 4-vector
$u_\emptyset^{\,\mu}$ for which the (barred) zero value,
{\srm X}=$\emptyset$,  of the chemical index will be
reserved, i.e. we shall set $n_\emptyset=s$,
$n_\emptyset^{\,\mu}=s^\mu$.

While the other (particle) currents may still either be conserved, in the
sense of having $\nabla_{\!\mu}n_{_{\rm X}}^{\,\mu}=0$ for certain values
of the chemical index {\srm X}, or else may have divergence
$\nabla_{\!\mu}n_{_{\rm X}}^{\,\mu}$ of unrestricted, positive or
negative, sign - corresponding to the possibilities of particle creation
or destruction -  the second law of thermodynamics stipulates that entropy
should never be destroyed, which means that in an entirely self contained
treatment we must always have
\be \nabla_{\!\mu}s^{\,\mu}\geq 0\, ,\label{2}\fe
i.e.  $\nabla_{\!\mu}n_\emptyset^{\,\mu}$ can never be negative. In some
contexts~\cite{CLS} it may however be convenient to work in terms of an open 
(i.e.not completely self contained) model in which, although not actually
destroyed, entropy is nevertheless effectively lost from the system by
some local heat removal mechanism - such as the URCA (neutrino -
antineutrino pair creation) process in a neutron star core -  in which
case the relevant remaining entropy current $s^\mu$ would not necessarily
have to respect the restriction (\ref{2}), but would be subject to the
modified inequality (\ref{24}) that is given below.

In the preceeding article~\cite{CCII} it was shown how, in a system
governed by a multifluid action variation principle, the relevant (kinetic,
internal, and gravitational) Lagrangian action contributions give rise to
corresponding variationally defined stress energy contributions that
combine to give a total $T_{_{\!\rm tot}\nu}^{\ \mu}$  which satisfies a
Noether type identity of the form
\be \nabla_{\!\mu}T_{_{\!\rm tot}\nu}^{\ \mu}=
{_\sum \atop ^{_{\rm X}}} f^{_{\rm X}}_{\,\mu}\, ,\label{4}\fe
in which, for each value of the chemical index ${\srm X}$, the 4-covector
$f^{_{\rm X}}_{\,\mu}$ denotes the variationally defined
non - gravitational force density acting on the correponding constituent.
The strictly conservative case considered in the preceeding work was
characterised by dynamical equations given, according to the variation
principle, just by the requirement that each of the separate force
densities $f^{_{\rm X}}_{\,\mu}$ should vanish. The purpose of the present
article is to extend the analysis to a more general category of dynamical
equations, whereby the force densities are not required to vanish but are
given by non-conservative contributions from dissipative mechanisms of
three different kinds, namely as viscosity, resistance against relative
motion, and transfusion between the various chemical constituents.

Although the suspension of the variation principle leaves a considerable
amount of lattitude in the way the various kinds of dissipative force
may be specified, the admissible forms of force law are considerably
restricted by the requirement of compatibility with the second
law of thermodynamics as embodied, for a self contained system,
in the inequality (\ref{2}). As in the preceding articles~\cite{CCI,CCII}
our work will be guided by previous experience~\cite{C89,C91,LSC98} with
analogous dissipative effects in a General Relativistic framework, which
(contrary to what is commonly supposed) is actually simpler
for many purposes, and particularly for the treatment of electromagnetic
effects, which are not included (except as possible external background
forces) in the present strictly Newtonian analysis.

\bigskip 
 {\bf 2. Viscous stress}
\medskip

The first of the dissipative mechanisms that we need to consider -
and the only one that will occur in a single constituent fluid model
- is that of {\it viscosity}, whose effect will be interpretable
in terms of a gross stress energy density tensor
\be T_{_{\!\rm gro}\nu}^{\ \mu}=T_{_{\!\rm tot}\nu}^{\ \mu}
+{_\sum\atop ^{_{\rm X}}} 
 \tau_{\,\nu}^{_{\rm X}\,\mu}, \label{5}\fe
in which the total $T_{_{\!\rm tot}\nu}^{\ \mu}$ provided by the
previously considered action contributions~\cite{CCII} is supplemented 
by further by viscous stress contributions $\tau_{\,\nu}^{_{\rm X}\,\mu}$
that are not obtained from the Lagrangian action but that are included
to allow for deviations (from what would otherwise be a local thermal
equilibrium state) due to space gradients of the corresponding flow 
vectors $u_{_{\rm X}}^{\,\mu}$. In accordance with what is suggested
by detailed microscopic analysis of dilute gas models~\cite{IsraelStewart79} 
it will be assumed that each such contribution has contravariant components 
\be \tau^{_{\rm X}\,\mu\nu}=\gamma^{\mu\rho} \tau_{\,\rho}^{_{\rm X}\,\nu}
\label{6}\fe
that are symmetric and purely spacelike, i.e.
\be \tau^{_{\rm X}\,\mu\nu}=\tau^{_{\rm X}\,\nu\mu}\, ,\hskip 1cm
\tau^{_{\rm X}\,\mu\nu}t_\nu=0\, ,\label{7}\fe
(where, as discussed in the preceeding work~\cite{CCI},
$ \gamma^{\mu\rho}$ is the degenerate Newtonian space metric
while $t_\mu$ is the preferred Newtonian time gradient) and it will also
be assumed that the mixed version is strictly spacelike with respect
to the corresponding fluid rest frame, i.e. for each value of the chemical
index {\srm X} we shall have
\be u_{_{\rm X}}^{\,\nu}\tau_{\,\nu}^{_{\rm X}\,\mu}=0\, .\label{8}\fe
It follows that it will be expressible in the form
\be\tau_{\,\nu}^{_{\rm X}\,\mu}=\gamma_{_{\rm X}\nu\rho}
\tau^{_{\rm X}\,\rho\mu}\, ,\label{12}\fe
where $\gamma_{_{\rm X}\nu\rho}$ is the positive indefinite
(rank 3) space metric tensor that would be determined
(in the manner described in the preceeding work~\cite{CCI}) by choosing
the  ether reference vector $e^\mu$ to coincide with the local
flow vector $u_{_{\rm X}}^{\,\nu}$, i.e. it is given by the defining
relations
\be \gamma_{_{\rm X}\nu\rho}\gamma^{\rho\mu}=
\gamma^{\,\mu}_{_{\rm X}\nu}\, ,\hskip 1 cm 
\gamma_{_{\rm X}\nu\rho}u_{_{\rm X}}^{\,\rho}=0\, ,\label{13}\fe
with
\be \gamma^{\,\mu}_{_{\rm X}\nu}=\delta_\nu^{\,\mu}-
u_{_{\rm X}}^{\,\mu}t_\nu\, .\label{14}\fe

In order to set up an appropriate category of models, we proceed
on the basis of the postulate that this gross stress energy tensor
should satisfy an energy momentum balance condition of the form
\be \nabla_{\!\mu}T_{_{\!\rm gro}\nu}^{\ \mu}=
{_\sum\atop ^{_{\rm X}}} f^{_{\rm X}}_{\!_{\rm ext}\,\mu}\, ,\label{9}\fe
in which the terms on the right will all vanish whenever we are dealing 
with a strictly self contained system, but in which the possibility
of external force density contributions $f^{_{\rm X}}_{_{\!\rm ext}\,\mu}$
is included to allow for cases when we are dealing with an open system 
involving effects such as neutrino emission or interaction with a long 
range electromagnetic field whose treatment within a model of the
non-relativistic kind studied here is prevented by the incompatibility 
of the necessary Lorentz and Galilean invariance requirements.

Using the specification (\ref{5}) in conjunction with the Noether
identity (\ref{4}) we see that the dynamical force balance requirement
(\ref{9}) will be expressible simply as
\be {_\sum\atop ^{_{\rm X}}} {\tilde f}{^{_{\rm X}}_{\,\mu}}=0\, 
,\label{10}\fe
where, for each constituent with label {\srm X}, the correspond
amalgamated force contribution is defined by 
\be {\tilde f}{^{_{\rm X}}_{\,\mu}}=f{^{_{\rm X}}_{\,\mu}}+\nabla_{\!\nu}
\tau_{\,\mu}^{_{\rm X}\,\nu}-f^{_{\rm X}}_{\!_{\rm ext}\,\mu}
\, .\label{11}\fe

\bigskip
{\bf 3. The thermodynamic positivity requirement.}
\medskip

It is now to be recalled that, according to our preceeding work,
each of the ordinary local constituent 4-force densities in the Noether
identity (\ref{4}) will be given as the sum of an acceleration
contribution and a (gauge dependent) 4-momentum transfer contribution
by the formula
\be f{^{_{\rm X}}_{\,\mu}}= \rlap{\,-} f{^{_{\rm X}}_{\,\mu}}
+\pi{^{_{\rm X}}_{\,\mu}}\nabla_{\!\nu}
n_{_{\rm X}}^{\,\nu}\, ,\label{15}\fe
in which $\pi{^{_{\rm X}}_{\,\mu}}$ is the relevant (gauge dependent) 
4-momentum covector that is defined in terms of the corresponding locally
defined material 4-momentum covector $\mu{^{_{\rm X}}_{\,\mu}}$
and the background gravitational field potential $\phi$ by
\be \pi{^{_{\rm X}}_{\,\mu}}=\mu{^{_{\rm X}}_{\,\mu}}- m^{_{\rm X}}
\phi t_\mu\, ,\label{16}\fe
where $m^{_{\rm X}}$ is the relevant particle rest mass parameter,
and the (gauge independent) acceleration contribution is specified
-- using a cross barred symbol --  by
\be  \rlap{\,-} f{^{_{\rm X}}_{\,\mu}}=2n_{_{\rm X}}^{\,\nu}\nabla_{\![\nu}
\pi^{_{\rm X}}{_{\!\mu]}}\, ,\label{15a}\fe which therefore satisfies
\be u_{_{\rm X}}^{\,\mu} \rlap{\,-} f{^{_{\rm X}}_{\,\mu}}=0 \, 
. \label{15b}\fe

In the particular case of the entropy current labelled by the value
{\srm X}=$\emptyset$, the relevant particle rest mass vanishes, and (as 
illustrated by the example of the historic Landau model~\cite{CK94}) 
the corresponding thermal momentum will be directly identifiable with
the local temperature covector that is obtained as the partial derivative 
of the material lagrangian density $\Lambda$ with respect to the entropy 
current, $\Theta_\mu=\partial\Lambda/\partial s^\mu$ (for fixed values 
of the other currents) and from which the temperature in the thermal rest 
frame is obtainable as 
\be\Theta=-u_\emptyset^{\,\mu}\Theta_\mu \, ,\label{17}\fe
 i.e. we shall simply have
\be m^\emptyset=0\, ,\hskip 1 cm \pi{^\emptyset_{\,\mu}}=\Theta_\mu
\, .\label{18}\fe
This means that, according to ({15}) the corresponding thermal 4-force 
density covector will be given just by
\be f{^\emptyset_{\,\mu}}=2s^{\,\nu}\nabla_{\![\nu}
\Theta_{\mu]}+\Theta{_{\mu}}\nabla_{\!\nu}
s^{\,\nu}\, ,\label{19}\fe
with the implication that its time component in the thermal rest
frame will be given simply by
\be u_\emptyset^{\,\nu} f{^\emptyset_{\,\nu}}=-\Theta\nabla_{\!\nu}
s^\nu\, ,\label{20}\fe
which will be negative (since the temperature $\Theta$ must always
be positive) by the second law requirement (\ref{2}) in any system
that is closed in the strong sense of being fully self contained,
so that the external forces $f^{_{\rm X}}_{\!_{\rm ext}\,\mu}$ on the 
right of (\ref{9}) will vanish. In the more general case of an open
system, in which there may be heat (i.e. thermal energy) loss at a 
rate (per unit volume in the thermal rest frame) given by
\be {\cal Q}=u_\emptyset^{\,\nu} f^\emptyset_{\!_{\rm ext}\,\nu} 
\, ,\label{23}\fe
due to mechanisms such as neutrino emission, the local formulation of the 
second law of thermodynamics will no longer take simple form (\ref{2}) but 
will be given by the more ubiquitously valid condition
\be {\cal Q} 
+\Theta \nabla_{\!\nu} s^\nu \geq 0\, .\label{24}\fe

\bigskip
{\bf 4. Transfusive dissipation}
\medskip

We use the term transfusion to designate processes by which particles 
of different species are converted into each other by various chemical 
or nuclear reaction processes, which we shall distinguish by capital 
Greek letters. The elementary process in some such reaction, with label 
$\Xi$ say, will involve the creation of a number $N_{_{\rm A}}^{_\Xi}$ say
of particles of the species with label {\srm A}, using the convention that 
the early Latin capital {\srm A} ranges over the same values as the late 
Latin index {\srm X} except for the zero value reserved for the entropy, 
i.e. it is subject to the restriction {\srm A}$\neq \emptyset$. (A simple 
example of such a process is the decomposition of a Helium nucleus into a  
pair of neutrons and a pair of protons, so that if we attribute the
labels 1,2,3 to neutrons, protons, and Helium nuclei respectively, this
reaction will be characterised by $N_{_{\rm 1}}^{_\Xi}=2$,
$N_{_{\rm 2}}^{_\Xi}=2$, $N_{_{\rm 3}}^{_\Xi}=-1$). In any such reaction,
the relevant particle creation numbers are restricted to satisfy the 
Newtonian mass conservation condition given by
\be {_\sum \atop ^{_{\rm A}}} N_{_{\rm A}}^{_\Xi}m^{_A}=0 .\label{30}\fe
By summing over the rates $r_{_\Xi}$ of the relevant reactions, the
ensuing particle creation rates are obtainable as
\be \nabla_{\!\nu}n_{_{\rm A}}^{\, \nu}={_\sum \atop ^{_{\Xi}}}r_{_\Xi} 
N_{_{\rm A}}^{_\Xi}\, ,\label{31}\fe
subject to the proviso that we are dealing with a system that is closed 
in the weak sense~\cite{Prigogine67}, meaning that there are no external 
losses or gains of the relevant particle species. Closure in the strong 
sense, meaning the condition that the model be entirely isolated in the 
sense of being fully self contained, would imply that all the external 
4-force density covectors $f^{_{\rm X}}_{\!_{\rm ext}\,\mu}$ in (\ref{9}) 
should vanish, whereas the weak closure condition adopted here 
corresponds merely to the requirement that the relevant material rest 
frame components should vanish, i.e.
\be f^{_{\rm A}}_{\!_{\rm ext}\,\nu}u_{_{\rm A}}^{\, \nu}=0
\, .\label{32}\fe
This restriction does not exclude the possibility of the kind of external 
force that might be exerted by interaction with a magnetic field, and 
since it does not apply to the special index value {\srm X}=$\emptyset$ 
labelling the entropy, it is also consistent with the possibility of 
a positive value ${\cal Q}>0$ of the heat loss rate (\ref{23}) due to a 
mechanism (such as the URCA process) of the kind that would necessitate 
replacement of the simple version (\ref{2}) of the second law of 
thermodynamics by  the more generally applicable version (\ref{24}).

In accordance with traditional usage~\cite{Prigogine67} in physical 
chemistry, it is convenient to work with a quantity of the kind for which 
De Donder introduced the term affinity, according to a specification
of the form
\be {\cal A}^{_\Xi}_{\{\epsilon\}}=-{_\sum \atop ^{_{\srm A\neq \emptyset}}}
N_{_{\rm A}}^{_\Xi}{\cal E}^{_{\rm A}}_{\{\epsilon\}}\, ,\label{34}\fe
in which the quantities ${\cal E}^{_{\rm A}}_{\{\epsilon\}}$
are the relevant energies per particle of the various species involved,
so that the affinity measure the net energy release in a elementary
process of the kind (characterised by the label $\Xi$) under consideration.

 A local static chemical (or nuclear) 
equilibrium state is one in which the relevant affinities vanish. When the 
deviations from such a state are not too large, it is naturally to be 
presumed that the reaction rates will be linearly dependent on the 
affinities according to a prescription of the form  
\be r_{_\Xi}={_\sum\atop ^{_{\Psi}}}\kappa_{_{\Xi\Psi}}
{\cal A}^{_\Psi}_{\{\epsilon\}}\, ,\label{35}\fe
in which the coefficients $\kappa_{_{\Xi\Psi}}$ form a symmetric matrix 
that must be positive in order to ensure the positivity of the energy 
release rate $\ssum r_{_\Xi}{\cal A}^{_\Xi}_{\{\epsilon\}}$. This implies 
that the particle creation rates will be given by the formula
\be \nabla_{\!\nu}n_{_{\rm A}}^{\, \nu}=-{_\sum\atop ^{_{\rm B}}}\Xi_{_{AB}}
{\cal E}^{_B}_{\{\epsilon\}} \, ,\label{36}\fe
in which $\Xi_{_{AB}}$ is a positive indefinite  matrix given by
\be \Xi_{_{AB}}={_\sum\atop ^{_{\Xi,\Psi}}} N_{_{\rm A}}^{_\Xi}
\kappa_{_{\Xi\Psi}} N_{_{\rm B}}^{_\Psi}\, .\label{37}\fe
This matrix is not positive definite, but only positive indefinite,
because it evidently has a null eigenvector provided by the set of 
particle masses, which will satisfy the condition
${_\sum\atop ^{_{\rm B}}}  \Xi_{_{AB}}m^{_B}=0$ by the mass 
conservation law (\ref{30}). Other such  null eigenvectors will 
be provided by other relevant charge number conservation laws. 
(The ordinary  mass conservation law is interpretable as the 
Newtonian limit of what is given in a relativistic theory by the
baryon conservation law.)

The meaning of the foregoing reasonning is unambiguous under conditions
of the kind most commonly considered in physical chemistry
in which there are no significant relative motions of the various
constituents. However in the more general circumstances we wish to
deal with here the meaning of the prescription (\ref{34}) will be
affected by the choice of reference systems in specification
of the energies ${\cal E}^{_{\rm A}}_{\{\epsilon\}}$.  The
most obvious possibility is to evaluate the energy in the thermal
rest frame, with respect to which the energy per particle will be
given simply by
\be {\cal E}^{_{\rm A}}_\emptyset=-\pi^{_A}_{\,\nu}
u_\emptyset^{\,\nu}\, .\label{34c}\fe
Such a specification does of course depend on the ether frame
used for the definition of the 4-momentum: it can be seen from
the analysis of the preceding article~\cite{CCI} that under
a transformation characterised by a Galilean boost velocity
vector $b^\mu=\gamma^{\mu\nu}\nabla_{\!\nu}\beta$  it
will be subject to a change given by the rule
\be{\cal E}^{_{\rm A}}_\emptyset\mapsto \breve
{\cal E}^{_{\rm A}}_\emptyset={\cal E}^{_{\rm A}}_\emptyset
+m^{_{\rm A}} u_\emptyset^{\,\nu}\nabla_{\,\nu}\beta-{_1\over ^2}
m^{_{\rm A}} b^2\, .\label{34d}\fe
However the mass conservation condition (\ref{30}) can be seen
to ensure that this gauge dependence will cancel out in the
corresponding thermal affinity, which is given by
\be {\cal A}^{_\Xi}_\emptyset=-{_\sum \atop ^{_{\srm A\neq \emptyset}}}
N_{_{\rm A}}^{_\Xi}{\cal E}^{_{\rm A}}_\emptyset
={_\sum \atop ^{_{\srm A\neq \emptyset}}}
N_{_{\rm A}}^{_\Xi}\pi^{_{\rm A}}_{\,\nu}u_\emptyset^{\,\nu}=
\breve{\cal A}^{_\Xi}_\emptyset
\, .\label{34a}\fe

Although it is independent of the choice of ether  frame, the
specification (\ref{34a}) of the thermal affinity  
${\cal A}^{_\Xi}_\emptyset$ will still be indeterminate in 
applications for which there is no well defined thermal reference
frame, a disadvantage that does not apply to what may be termed
the natural affinity, which is given in terms of the corresponding
natural energies $\chi^{_{\rm A}}_\natural$ by
\be {\cal A}^{_\Xi}_\natural=-{_\sum \atop ^{_{\srm A\neq \emptyset}}}
N_{_{\rm A}}^{_\Xi}\chi^{_{\rm A}}_\natural 
=\breve{\cal A}^{_\Xi}_\natural
\, .\label{34e}\fe
The natural energy is  specified for each constituent as the corresponding
chemical potential as measured with respect to its own local rest frame, 
which means that it will be given  by 
\be \chi^{_{\rm A}}_\natural=-u_{_{\rm A}}^{\,\mu}\chi^{_{\rm A}}_{\,\mu}
=-u_{_{\rm A}}^{\,\mu}\pi^{_{\rm A}}_{\,\mu}+{_1\over^2}m^{_{\rm A}}
v_{_{\rm A}}^{\, 2}-m^{_{\rm A}}\phi\, ,\label{34f}\fe
in which the gauge dependence of the three separate terms on the right
cancels out, to give 
\be \breve\chi^{_{\rm A}}_\natural
=\chi^{_{\rm A}}_\natural \, .\fe

In the absence, at this stage, of any clear idea of which, if any
of these two (thermal  and natural) alternatives may be most
appropriate for general purposes, we shall proceed in terms of a 
compromise using a mixed energy  ${\cal E}^{_{\rm A}}_{\{\epsilon\}}$
that is defined in terms of a parameter $\epsilon$ by
\be {\cal E}^{_{\rm A}}_{\{\epsilon\}}=(1-\epsilon)
{\cal E}^{_{\rm A}}_\emptyset  +\epsilon\, \chi^{_{\rm A}}_\natural
\, ,\label{39}\fe
which means that it will transform according to the rule
 \be {\cal E}^{_{\rm A}}_{\{\epsilon\}}\mapsto\breve
{\cal E}^{_{\rm A}}_{\{\epsilon\}}={\cal E}^{_{\rm A}}_{\{\epsilon\}}
+m^{_{\rm A}}(1-\epsilon\,) \big( u_\emptyset^{\,\nu}
\nabla_{\,\nu}\beta-\frac{_1}{^2} b^2\big) \, ,\label{39a}\fe 
so that as before, in consequence of the mass conservation condition
(\ref{30}), the corresponding affinity (\ref{31}) will be invariant,
\be  \breve{\cal A}^{_\Xi}_{\{\epsilon\}}={\cal A}^{_\Xi}_{\{\epsilon\}}
\, .\label{39b}\fe
In this paramatrised weighting scheme the special thermal and natural
cases are given respectively by
\be {\cal A}^{_\Xi}_{\{0\}}={\cal A}^{_\Xi}_\emptyset\, ,\hskip
1 cm {\cal A}^{_\Xi}_{\{1\}}={\cal A}^{_\Xi}_\natural\, ,\label{39c}\fe
and the general case we shall have 
\be {\cal A}^{_\Xi}_{\{\epsilon\}}=(1-\epsilon\,){\cal A}^{_\Xi}_\emptyset
+\epsilon\,{\cal A}^{_\Xi}_\natural\, .\label{39d}\fe
 
When the chemical reaction rates and the relative velocities are
sufficiently small it will not matter what value is chosen for the 
parameter $\epsilon$, a consideration that presumably accounts for the 
lack of attention to this issue in the standard literature on 
non-equilibrium thermodynamics ~\cite{Prigogine67}. However if the 
chemical reaction rates or the relative velocities are too large, the
distinction may become important and in such a case the question of 
what would be most physically realistic would ultimately need to be 
decided on the basis of a microscopic analysis of a kind that does not 
yet seem to have been sufficiently developed. It might turn out that for
an accurate out of equilibrium thermodynamic description it would be 
most appropriate to use something more complicated than the kind of 
weighted mean adopted here. It is to be remarked that an affinity of the 
natural kind, as characterised in our present notation by the weighting 
ansatz $\epsilon=1$,  is what was implicitly used in earlier relativistic 
work~\cite{C89,C91}, whereas use of an affinity of the thermal kind, as 
characterised by  $\epsilon=0$, was implicit in more recent and specialised
relativistic work~\cite{LSC98}. It will be found below that the latter is 
more satisfactory for applications involving superconductivity.

\bigskip
{\bf 5. Viscous dissipation}
\medskip

In a systematic approach to the construction of phenomenological 
dissipation laws that are consistent with the thermodynamical inequality 
(\ref{24}), the first step is to evaluate its left hand side by contraction 
of the 4-force balance equation (\ref{10}) with the thermal rest frame unit 
vector $u_\emptyset^{\, \nu}$ so as to obtain an identity of the form
\be {_\sum \atop ^{_{\rm X}}}u_{_{\rm X}}^{\,\mu}
 f^{_{\rm X}}_{\!_{\rm ext}\,\mu}={_\sum \atop ^{_{\rm X}}}
\Big(u_\emptyset^{\,\mu}\pi^{_{\rm X}}_{\,\mu}\nabla_{\!\nu}
n_{_{\rm X}}^{\,\nu}+u_{_{\rm X}}^{\,\mu}
\nabla_{\!\nu}\tau_{\,\nu}^{_{\rm X}\,\mu}- v_{_{\rm X}\emptyset}^\mu\,
\tilde {\!\rlap{\,-}f}{^{_{\rm X}}_{\,\mu}} \Big)\, ,\label{40}\fe
where $v_{_{\rm X}\emptyset}^\mu$ is the (purely spacelike) velocity 
difference between the particular unit flow 4-vector $u_{_{\rm X}}^{\,\mu}$ 
and the thermal rest frame unit 4-vector  $u_\emptyset^{\,\mu}$, i.e.
\be v_{_{\rm X}\emptyset}^{\,\mu}=u_{_{\rm X}}^{\,\mu} 
-u_\emptyset^{\,\mu}\, ,\label{41}\fe
and $\,\tilde{\!\rlap{\,-}f} {^{_{\rm X}}_{\,\mu}}$ is the gauge independent
force density contribution given by
\be \,\tilde{\! \rlap{\,-}f} {^{_{\rm X}}_{\,\mu}}=
\tilde f {^{_{\rm X}}_{\,\mu}}-\pi^{_{\rm X}}_{\,\mu}
\nabla_{\!\nu} n_{_{\rm X}}^{\,\nu}= \rlap{\,-}f {^{_{\rm X}}_{\,\mu}}
+\nabla_{\!\nu}\tau_{\,\mu}^{_{\rm X}\,\nu}- 
f^{_{\rm X}}_{\!_{\rm ext}\,\mu}\, .\label{41b}\fe

Let us now generalise this specification to a parameter dependent
force density contribution given by an expression of the analogous form
\be \,\tilde{f} {^{_{\rm X}}_{\!\{\epsilon\}\mu}}=
f {^{_{\rm X}}_{\!\{\epsilon\}\mu}}+\nabla_{\!\nu}
\tau_{\,\mu}^{_{\rm X}}{^\nu}- f^{_{\rm X}}_{\!_{\rm ext}\,\mu}
\, ,\label{41c}\fe
in terms of a combination given for {\srm A}$\neq\emptyset$,  as a function  
of the weighting parameter $\epsilon$ of the preceding section, by the formula
\be f{^{_{\rm A}}_{\!\{\epsilon\}\mu}}=
 f{^{_{\rm A}}_{\,\mu}}-\Big( \pi^{_{\rm A}}_{\, \mu}-\epsilon\,\big(
 \chi^{_{\rm A}}_{\, \mu}+\frac{_1}{^2} m^{_{\rm A}}
v_{_{\rm A}\emptyset\,\mu}\big)\Big) \nabla_{\!\nu}n_{_{\rm A}}^{\,\nu}
\, ,\label{41d}\fe
so that in particular we shall have
\be f{^{_{\rm A}}_{\!\{0\}\mu}}={\! \rlap{\,-}f} {^{_{\rm A}}_{\,\mu}}
\, .\label{41e}\fe
We complete the specification for the thermal case with a formula of the
rather different form
\be f{^\emptyset_{\!\{\epsilon\}\mu}}=f{^\emptyset_{\,\mu}}
+{_\sum \atop ^{_{\rm A}}}
\Big( \pi^{_{\rm A}}_{\, \mu}-\epsilon\,\big(
 \chi^{_{\rm A}}_{\, \mu}+\frac{_1}{^2} m^{_{\rm X}}
v_{_{\rm A}\emptyset\,\mu}\big)\Big) \nabla_{\!\nu}n_{_{\rm A}}^{\,\nu}
\, ,\label{41h}\fe
in order to obtain a sum over all constituents that is the same as
for the ordinary forces:
\be {_\sum \atop ^{_{\rm X}}}f{^{_{\rm A}}_{\!\{\epsilon\}\mu}}=
{_\sum \atop ^{_{\rm X}}}f{^{_{\rm X}}_{\,\mu}}\, .\label{41g}\fe

This specification has been set up in such a way that, unlike the 
original (canonically defined) 4-force vectors $f{^{_{\rm X}}_{\,\mu}}$, 
which are frame dependent unless the number currents are separately
conserved, the adjusted 4-force densities given by (\ref{41d}) and 
(\ref{41h}) will have space projected parts that are always unaffected by 
Galilean (and even Milne) transformations, i.e. they will satisfy the 
invariance conditions
\be \gamma^{\mu\nu}\breve 
 f{^{_{\rm X}}_{\!\{\epsilon\}\nu}}=\gamma^{\mu\nu}
 f{^{_{\rm X}}_{\!\{\epsilon\}\nu}}\, .\label{41f}\fe

The motivation for the introduction of the parametrically adjusted
force densities given by the rather elaborately contrived definition
(\ref{41d}) is that it enables us to obtain a particularly simple and 
evocative expression for the entropy term in (\ref{40}). Using the total 
mass conservation condition that is obtainable from (\ref{30}) in the form
\be{_\sum \atop ^{_{\rm A\neq\emptyset}}}m^{_{\rm A}}\nabla_{\!\nu}
n_{_{\rm A}}^{\,\nu}=0 \, ,\label{43}\fe
in conjunction with the restrictions (\ref{8}) and (\ref{32})
it can be seen to follow from (\ref{40}) that will be given --
for any chosen value of the weighting parameter $\epsilon$ --
by
\be {\cal Q} +\Theta 
\nabla_{\!\mu} s^\mu=-{_\sum \atop ^{_{\rm A}}}
{\cal E}^{_A}_{\!\{\epsilon\}} \nabla_{\!\nu}n_{_{\rm A}}^{\,\nu}
-{_\sum \atop ^{_{\rm X}}}
\tau_{\,\nu}^{_{\rm X}\,\mu}\nabla_{\!\mu}u_{_{\rm X}}^{\,\nu}
-{_\sum \atop ^{^{\rm A}}} v_{_{\rm A}\emptyset}^\mu 
\,\tilde{f} {^{_{\rm A}}_{\!\{\epsilon\}\,\mu}}
 \, .\label{42}\fe

In view of the second law requirement (\ref{24}) to the effect that
the left hand side of (\ref{42}) should be positive, the choice of
admissible dissipation laws will be restricted by the condition that
it should be such as to ensure that the sum of the terms on the 
right of (\ref{42}) should also be positive. Although many other 
(generally more complicted) ways involving various kinds of cross 
coupling are conceivable, it will be adequate for most 
purposes to do this in the most obvious way by ensuring that each of 
the three sums on the right of (\ref{42}) is separately positive. 

In so far as the first of these terms is concerned, this desideratum 
of positivity is already satisfied by the ansatz of the previous section, 
which -- for the chosen value of $\epsilon$ -- gives an expression of the form
\be -{_\sum \atop ^{_{\rm A\neq\emptyset}}}
{\cal E}^{_A}_{\!\{\epsilon\}} \nabla_{\!\nu}n_{_{\rm A}}^{\,\nu}=
{_\sum\atop ^{_{\Xi,\Psi}}}{\cal A}^{_\Xi}_{\!\{\epsilon\}} \kappa_{_{\Xi\Psi}}
{\cal A}^{_\Psi}_{\!\{\epsilon\}} \, ,\label{44}\fe
whose positivity is evidently ensured by the condition that the
transfusion matrix $\kappa_{_{\Xi\Psi}}$ should be positive definite.

To deal with the second term, we exploit the possibility of rewriting  the 
negative of the contribution of each constituent as
\be\tau_{\,\nu}^{_{\rm X}\,\mu}\nabla_{\!\mu}u_{_{\rm X}}^{\,\nu}
=\gamma_{\mu\rho}\tau^{_{\rm X}\,\rho\nu}
\gamma_{\nu\sigma}\,\theta_{_{\rm X}}^{\,\mu\sigma}\, ,\label{50}\fe
where $\theta_{_{\rm X}}^{\,\mu\sigma}$ is the symmetric spacelike
expansion rate tensor given  by
\be \theta_{_{\rm X}}^{\,\mu\sigma}=\gamma^{\nu(\mu}\nabla_{\!\nu}
u_{_{\rm X}}^{\,\sigma)}\, ,\hskip 1 cm \theta_{_{\rm X}}^{\,\mu\sigma}
t_\sigma=0\, .\label{51}\fe
This tensor is decomposible in a well defined (Galilean and even Milne
gauge independent) manner into tensorially irreducible parts in the form
\be \theta_{_{\rm X}}^{\,\mu\nu}= \sigma_{_{\rm X}}^{\,\mu\nu}
+{_1\over^3}\theta_{_{\rm X}}\gamma^{\mu\nu}\, ,\label{52}\fe
where $\sigma_{_{\rm X}}^{\,\mu\nu}$ is the trace free shear rate tensor 
and $\theta_{_{\rm X}}$ is the scalar expansion rate, as characterised by
\be \gamma_{\mu\nu}\sigma^{\mu\nu}=0\, ,\hskip 1 cm
\theta_{_{\rm X}}=\gamma_{\mu\nu}\theta_{_{\rm X}}^{\,\mu\nu}\, .\label{53}\fe
This enables us to write the negative of viscosity term in (42) as
\be {_\sum\atop ^{_{\rm X}}} 
\tau_{\,\nu}^{_{\rm X}\,\mu}\nabla_{\!\mu}u_{_{\rm X}}^{\,\nu}
 ={_\sum \atop ^{_{\rm X}}}\gamma_{\mu\rho}\gamma_{\nu\sigma}
\tau^{_{\rm X}\,\mu\nu}\sigma_{_{\rm X}}^{\,\rho\sigma}+{_1\over^3}
{_\sum\atop ^{_{\rm X}}}\gamma_{\mu\nu}\tau^{_{\rm X}\,\mu\nu}
\theta_{_{\rm X}}\, .\label{54}\fe

In order to ensure that this total is negative as required, the obvious
generalisation of the ansatz that is familiar in the case of a single
constituent fluid is to postulate that the relevant stress contributions
are given by an expression of the form
 \be\tau^{_{\rm X}\,\mu\nu}=-2{_\sum\atop ^{_{\rm Y}}}
\eta^{_{\rm XY}}\sigma_{_{\rm Y}}^{\,\mu\nu}
-{_\sum\atop ^{_{\rm Y}}}
\zeta^{_{\rm XY}}\theta_{_{\rm Y}}\gamma^{\,\mu\nu}\, ,\label{55}\fe
where $\eta^{_{\rm XY}}$ is a positive definite definite or indefinite
but in any case (by the Onsager principle) symmetric matrix of shear 
viscosity coefficients, and $\zeta^{_{\rm XY}}$ is a similarly symmetric 
positive definite or indefinite matrix of bulk viscosity coefficients.

\bigskip
{\bf 6. Ordinary resistive dissipation}
\medskip

To complete the determination of the dynamical equations of motion
it remains to specify the space components of the force on each
constituent. The most obvious way of doing this in such a way as to
ensure consistency with the total force balance condition
(\ref{10}) is to take them to consist of sums of pairwise interaction
contributions in the form
\be \gamma^{\mu\nu} \,\tilde{f} {^{_{\rm X}}_{\!\{\epsilon\}\,\nu}}
={_\sum\atop ^{_{\rm Y}}} f^{_{\rm XY}}{^\mu}\, ,\label{60a}\fe
subject to the conditions
\be f^{_{\rm XY}\,\mu} =-f^{_{\rm YX}\,\mu}\, ,\hskip 1 cm
 t_\mu f^{_{\rm XY}\,\mu}=0\, .\label{60b}\fe
It is possible to conceive situations in which a more
elaborate construction procedure might be needed, but we shall
not envisage such complications here.

Proceding on the basis of the ansatz (\ref{60a}) we must now 
consider how the admissible forms of the two-constituent interaction
force densities $f^{_{\rm XY}\,\mu}$ are restricted by the
second law of thermodynamics. As we have already chosen rules that ensure 
the positivity of the first two terms on the right of (\ref{42}), this 
restriction will amount just to the requirement of positivity of the final 
term, which will be given by
\be-{_\sum\atop ^{_{\rm X}}} v_{_{\rm X}\emptyset}^\mu 
\,\tilde{f} {^{_{\rm X}}_{\!\{\epsilon\}\,\mu}}
={_\sum\atop ^{_{\rm X,Y}}} v_{_{\rm X}\emptyset\,\mu}
f^{_{\rm YX}\,\mu}=\frac{_1}{^2}
{_\sum\atop ^{_{\rm X,Y}}} v_{_{\rm XY}\,\mu}
f^{_{\rm YX}\,\mu}
\, .\label{60}\fe
in which the (gauge invariant) relative velocity  $v_{_{\rm XY}}^\mu$ 
and the corresponding (gauge dependent) covector $v_{_{\rm XY}\,\mu}$
are defined by
\be v_{_{\rm XY}}^\mu= v_{_{\rm X}\emptyset}^\mu -v_{_{\rm Y}\emptyset}^\mu
= u_{_{\rm X}}^\mu -u_{_{\rm Y}}^\mu\, ,
\hskip 1 cm v_{_{\rm XY}\,\mu}=\gamma_{\mu\nu}v_{_{\rm XY}}^\nu\, .
\label{62}\fe

The obvious way to fulfil this requirement is to suppose that the forces 
are due just to resistivity of the ordinary kind, which means that
for each pair of distinct constituent label values {\srm}$\neq${\srm Y} 
the corresponding contribution will be given by an ordinary positive 
resistivity coefficient $Z^{_{\rm XY}}=Z^{_{\rm YX}}\geq 0$ according 
to the specification
\be f^{_{\rm YX}\,\mu}=Z^{_{\rm XY}}v_{_{\rm XY}}^\mu\, .\label{63}\fe

It follows that the third term on the right of (\ref {42}) will be given 
by 
\be -{_\sum\atop ^{_{\rm A}}} v_{_{\rm A}\emptyset}^\mu 
\,\tilde{f} {^{_{\rm A}}_{\!\{\epsilon\}\,\mu}}=
{_1\over^2}{_\sum\atop ^{_{\rm X,Y}}}
Z^{_{\rm XY}}v_{_{\rm XY}}^\mu \gamma_{\mu\nu}  v_{_{\rm XY}}^\nu
\, ,\label{66}\fe
which shows that it does indeed satisfy the required positivity 
condition.

Having thus obtained an appropriate resistivity formula (\ref{60a})
for the gauge invariant force density components 
$\gamma^{\mu\nu}\tilde{f} {^{_{\rm X}}_{\!\{\epsilon\}\,\nu}}$
 we can imediately use the defining relation 
\be \tilde f{^{_{\rm A}}_{\,\mu}}=\tilde f{^{_{\rm A}}_{\!\{\epsilon\}\mu}}
-\Big( \pi^{_{\rm A}}_{\, \mu}-\epsilon\,\big(
 \chi^{_{\rm A}}_{\, \mu}+\frac{_1}{^2} m^{_{\rm A}}
v_{_{\rm A}\emptyset\,\mu}\big)\Big) 
\nabla_{\!\nu}n_{_{\rm A}}^{\,\nu}\, ,\label{67a}\fe
to provide a corresponding formula for the original unadjusted
(gauge dependent) material force density components, which will
be given  by
\be\gamma^{\mu\nu} \tilde f{^{_{\rm A}}_{\,\nu}}=
{_\sum\atop ^{_{\rm Y}}} Z^{_{\rm AY}} v_{_{\rm YA}}^\mu
-\Big( \gamma^{\mu\nu}\pi^{_{\rm A}}_{\, \nu}-\epsilon\,\big(\gamma^{\mu\nu}
 \chi^{_{\rm A}}_{\, \nu}+\frac{_1}{^2} m^{_{\rm A}}
v_{_{\rm A}\emptyset}^{\ \mu}\big)\Big) 
\nabla_{\!\nu}n_{_{\rm A}}^{\,\nu}\, .\label{67}\fe

\bigskip
{\bf 6. Superfluid drag dissipation}
\medskip

The simple kind of resistive dissipation mechanism described in the
previous section will not be operational  in the case of
 a constituent that is superfluid. To deal with such cases, let us
use indices {\srm I,J} that range over the values (if any) of {\srm X}
referring to a superfluid constituent, while using indices {\srm C,D}
that range over the values of {\srm X} referring
to the remaining -- normal -- constituents. 

On a sufficiently small -- mesoscopic -- scale, superfluidity can
be dealt with directly in terms of the kind of model set up in the
preceding section by imposing the following conditions, of which
the first and most obvious is simply the requirement that the 
relevant superfluid viscosity and resistivity coefficients should 
vanish, i.e.
 \be \eta^{_{\rm IX}}=0\, ,\hskip 1 cm
\zeta^{_{\rm IX}}\, ,\hskip 1 cm Z^{_{\rm IX}}=0\, ,\label{101}\fe
which implies, by (\ref{55}) that
that the corresponding viscous tension contributions
$\tau^{_{\rm I}\,\mu\nu}$ will vanish. The next condition is that
that the superfluid constituents are not directly subject to any 
external force, 
\be f^{_{\rm I}}_{\!_{\rm ext}\,\mu}=0,\label{102}\fe
so that we shall be able to make the identifications
\be \tilde f^{_{\rm I}}_{\,\mu}= f^{_{\rm I}}_{\,\mu}\, .\label{102a}\fe
The final condition that needs to be imposed is that we should 
be able to make the identification
\be \tilde{f} {^{_{\rm I}}_{\!\{\epsilon\}\,\mu}}=
{f} {^{_{\rm I}}_{\!\{0\}\,\mu}}\, ,\label{103a}\fe
where (by definition)
\be {f} {^{_{\rm I}}_{\!\{0\}\,\mu}} =n_{_{\rm I}}^{\,\nu}
\varpi^{_{\rm I}}_{\,\nu\mu} \, ,\label{103b}\fe
which will follow from (\ref{102a}) subject to the requirement that
we adopt the thermal affinity ansatz,
\be \epsilon=0 \, ,\label{103}\fe
or else that the relevant chemical reaction rates are set to zero so
that the superfluid particle creation rates $\nabla_{\!\nu}
n_{_{\rm I}}^{\,\nu}$ all vanish, in which case the value of
$\epsilon$ will not matter. When the conditions (\ref{101}), (\ref{102})
and (\ref{103}) are all satisfied it can be seen that the equations of
motion set up in the preceeding sections will be consistent with the
restraint to the effect that the relevant superfluid vorticities
\be\varpi^{_{\rm I}}_{\,\mu\nu}=2\nabla_{\![\mu}\pi^{_{\rm I}}_{\,\nu]}
\, ,\label {104}\fe
should vanish, as necessary for the existence of corresponding local
superfluid phase scalars $\varphi^{_{\rm I}}$ such that
$N_{_{\rm I}}\pi^{_{\rm I}}_{\,\nu}=\hbar\nabla_{\!\nu}\varphi^{_{\rm I}}$
where $N_{_{\rm I}}$ is the number (2 for the usual
case of a Cooper type pair) of constituent particle in a boson of
the superfluid condensate under consideration.

For application on larger scales such a mesoscopic description is
inadequate, and must be replaced by a macroscopic description that
allows non vanishing superfluid vorticities $\varpi^{_{\rm I}}_{\,\mu\nu}$,
which are interpretable a representing the average effect of a fibration
by microscopic vortex tubes on which the mesoscopic irrotationality
condition breaks down. This interpretation means that although it does
not have to vanish, the macroscopic vorticity 2-form of a superfluid 
constituent can not have an arbitrary algebraic form but must satisfy the 
degeneracy condition
\be\varpi^{_{\rm I}}_{\,[\mu\nu}\varpi^{_{\rm I}}_{\,\rho]\sigma}=0
\, ,\label{105}\fe
(which means that its matrix rank is not four as in the generic case
but two, since its antisymmetry excludes the possibility of an odd valued
rank) in order for the null eigenvectors that generate the vorticity
flux two surfaces to exist. As remarked in the context of the analogous
relativistic problem~\cite{LSC98},  the only obvious way of setting up a
force law that satisfies this condition is to postulate that it should
have the form
\be {f} {^{_{\rm I}}_{\!\{0\}\,\mu}}=n_{_{\rm I}}\varpi^{_{\rm I}}_{\,\mu\nu}
V_{_{\rm I }}^{\,\nu}\, ,\label{106}\fe
in which $V_{_{\rm I }}^{\,\nu}$ is some vector such that
the combination  $u_{_{\rm I }}^{\,\nu}+V_{_{\rm I }}^{\,\nu}$
is one of the null eigenvectors generating the flux surfaces
of the vorticity flux $\varpi^{_{\rm I}}_{\,\mu\nu}$, i.e. such that
$\varpi^{_{\rm I}}_{\,\mu\nu}(u_{_{\rm I }}^{\,\nu}+V_{_{\rm I }}^{\,\nu})=0$.

The superfluidity ansatz  (\ref{106}) can be applied within  the
framework set up in the preceding section by supposing that 
$V_{_{\rm I }}^{\,\nu}$ is decomposible as a sum of separate contributions
$V_{_{\rm I}}^{_{\rm C}\,\nu}$ from the various non-superfluid contributions,
in the form 
\be V_{_{\rm I }}^{\,\nu}= {_\sum\atop ^{_{\rm C}}}
V_{_{\rm I}}^{_{\rm C}\,\nu}\, ,\label{107}\fe
so that the corresponding space projected force contributions in the 
decomposition (\ref{60a}) can be taken to be given by
\be f^{_{\rm IC}\,\mu}=n_{_{\rm I}}\gamma^{\mu\nu}
\varpi^{_{\rm I}}_{\,\nu\rho}
V_{_{\rm I}}^{_{\rm C}\,\rho}\, .\label{107a}\fe

As in the preceding section, we now need to find a procedure for choosing
the force contributions in such a way as to ensure the positivity
of the final term on the right of (\ref{42}), which will now be given 
as a combination of normal and superfluid contributions in the form
\be-{_\sum\atop ^{_{\rm X}}} v_{_{\rm X}\emptyset}^\mu 
\,\tilde{f} {^{_{\rm X}}_{\!\{0\}\,\mu}}
=\frac{_1}{^2}{_\sum\atop ^{_{\rm C,D}}} v_{_{\rm CD}\,\mu}
f^{_{\rm DC}\,\mu}+{_\sum\atop ^{_{\rm C,I}}} v_{_{\rm CI}\,\mu}
f^{_{\rm IC}\,\mu}\, .\label{108}\fe
As before, we can deal with the normal part by taking the relevant
force contributions to be given for {\srm C}$\neq${\srm D} by a set of 
normal resistivity coefficients $Z^{_{\rm CD}}=Z^{_{\rm CD}}\geq 0$
according to the simple specification
\be f^{_{\rm DC}\,\mu}=Z^{_{\rm CD}}v_{_{\rm CD}}^\mu\, ,\label{109}\fe
which will automatically take care of the positivity of the first
term on the right of (\ref{108}),
but the superfluid contributions
$  f^{_{\rm IC}\,\mu}$ will need to be handled in a different manner.

The remaining positivity requirement that still needs to be satisfied, is
that of the second term on the right of (\ref{108}), which can be rewritten
as
\be {_\sum\atop ^{_{\rm C,I}}} v_{_{\rm CI}\,\mu} f^{_{\rm IC}\,\mu}
={_\sum\atop ^{_{\rm C,I}}}  v_{_{\rm CI}}^{\,\mu}\,n_{_{\rm I}}
\varpi^{_{\rm I}}_{\,\mu\nu} V_{_{\rm I}}^{_{\rm C}\,\nu}=
{_\sum\atop ^{_{\rm C,I}}}  u_{_{\rm C}}^{\,\mu}\, n_{_{\rm I}}
\varpi^{_{\rm I}}_{\,\mu\nu} V_{_{\rm I}}^{_{\rm C}\,\nu}
\, ,\label{111}\fe
where the last step is obtained by substituting (\ref{107}) in
(\ref{106}) and using the identity $u_{_{\rm I}}^{\,\mu}
 {f} {^{_{\rm I}}_{\!\{0\}\,\mu}}=0$. We now procede in a manner
analogous to that by which the ordinary resistivity forces were
introduced above, which means that we ensure the positivity of the
separate terms in the sum on the right of (\ref{111}) by taking each
vector $V_{_{\rm I}}^{_{\rm C}\,\mu}$ to be given by an expression
of the form
\be V_{_{\rm I}}^{_{\rm C}\,\mu}=- c_{_{\rm I}}^{\,_{\rm C}}
(w^{_{\rm I}})^{-1}\gamma^{\mu\nu}\varpi^{_{\rm I}}_{\,\nu\rho}
u_{_{\rm C}}^{\,\rho}\, ,\label{112}\fe
where $c_{_{\rm I}}^{\,_{\rm C}}$ is a positive drag coefficient and we
have included a positive normalisation factor given by the magnitude
$w^{_{\rm I}}\geq 0$ of the (gauge independent) space projected
vorticity vector
\be w^{_{\rm I}\,\mu}=\frac{_1}{^2}\varepsilon^{\mu\nu\rho}
\varpi^{_{\rm I}}_{\,\nu\rho}\, ,\label{113b}\fe
according to the specification
\be (w^{_{\rm I}})^2=\frac{_1}{^2}\gamma^{\mu\nu}
\gamma^{\rho\sigma}\varpi^{_{\rm I}}_{\,\mu\rho}
\varpi^{_{\rm I}}_{\,\nu\sigma}=\gamma_{\mu\nu}w^{_{\rm I}\,\mu}
w^{_{\rm I}\,\nu}\, .\label{113}\fe

The prescription (\ref{113}) is evidently equivalent to the
adoption of a force law of the form
\be f^{_{\rm IC}\,\mu}=c_{_{\rm I}}^{\,_{\rm C}}
n_{_{\rm I}}w^{_{\rm I}}\perp^{\!_{\rm I}\mu}_{\ \nu}
u_{_{\rm C}}^{\,\nu}\, ,\label{114}\fe
where
\be \perp^{\!_{\rm I}\mu}_{\ \nu}=  (w^{_{\rm I}})^{-2}
\gamma^{\mu\rho}\gamma^{\sigma\tau}\varpi^{_{\rm I}}_{\,\tau\rho}
\varpi^{_{\rm I}}_{\,\sigma\nu}\, .\label{115}\fe
Adding up the resulting contributions we finally obtain
\be \gamma^{\mu\nu}{f} {^{_{\rm I}}_{\!\{0\}\,\nu}}=n_{_{\rm I}}
\gamma^{\mu\nu}\varpi^{_{\rm I}}_{\,\nu\rho}
V_{_{\rm I }}^{\,\rho}\, ,\label{106b}\fe
which (by the identity  $u_{_{\rm I}}^{\,\nu}
{f} {^{_{\rm I}}_{\!\{0\}\,\nu}}=0$) automatically provides
the required result (\ref{106}) with
\be V_{_{\rm I }}^{\,\mu}=- (w^{_{\rm I}})^{-1}\gamma^{\mu\nu}
\varpi^{_{\rm I}}_{\,\nu\rho}{_\sum\atop ^{_{\rm C}}}
c_{_{\rm I}}^{\,_{\rm C}}u_{_{\rm C}}^{\,\rho}\, .\label{112b}\fe

To relate this 4-dimensionally covariant formulation to
the traditional Newtonian terminology using a 3+1 decomposition
based on some particular choice of ether frame vector $e^\mu$,
it is useful to introduce the (frame dependent) vorticity surface generating 
unit 4-vector $\check u_{_{\rm I}}^\mu$ and its associated 3-velocity
vector $\check v_{_{\rm I}}^\mu=\check u_{_{\rm I}}^\mu-e^\mu$
by the defining conditions
\be \varpi^{_{\rm I}}_{\,\mu\nu}\check u_{_{\rm I}}^\nu=0\, ,\hskip
1 cm \gamma_{\mu\nu} w^{_{\rm I}\,\mu}\check v_{_{\rm I}}^\nu=0
\, .\label{116}\fe
In terms of such a vorticity flux velocity vector, the
 degenerate vorticity 2-form $\varpi^{_{\rm I}}_{\,\mu\nu}$
will be expressible as
\be \varpi^{_{\rm I}}_{\,\mu\nu}=\big(\varepsilon_{\mu\nu\rho}
 + 2 t_{[\mu}\varepsilon_{\nu]\sigma\rho}
\check v_{_{\rm I}}^\sigma\big) w^{_{\rm I}\,\rho}\, ,\label{117}\fe
and the (rank 2) projection tensor in (\ref{114}) will be
given by 
\be \perp^{\!_{\rm I}\mu}_{\ \nu}=\gamma^\mu_{\,\nu}
-\check v_{_{\rm I}}^{\,\mu}t_\nu-(w^{_{\rm I}})^{-2}
w^{_{\rm I}\,\mu}w^{_{\rm I}}_{\,\nu}\, .\label{118}\fe
It can thus be seen that the projected velocity on the right
of (\ref{114}) will be given by
\be \perp^{\!_{\rm I}\mu}_{\ \nu} u_{_{\rm C}}^{\,\nu}=
 v_{_{\rm C}}^{\,\mu}
-(w^{_{\rm I}})^{-2}
w^{_{\rm I}\,\mu}w^{_{\rm I}\,\nu} v_{_{\rm C}\,\nu}
-\check v_{_{\rm I}}^{\,\mu}\, .\label{119}\fe

\bigskip
{\bf 7. The limit cases of convection and pinning}
\medskip

If some resistivity coefficient, $Z^{_{\rm XY}}$ is very large,
the corresponding velocity difference will tend to be
very small $v_{_{\rm YX}}^\mu$. In such a case it will often
be convenient to a use a simplified dynamical treatment based
on the relevant {\it convection ansatz} to the effect that
the velocity difference in question should actually vanish.
To deal with this kind of convective limit, i.e.  a case
in which the number of independent normal velocities is smaller that
the number of independent chemical constituents, it is convenient
to introduce a new kind of index $\langle {\srm U}\rangle $, distinguished
by surrounding angle brackets,  to label the independent velocities and the 
corresponding comoving subsets of constituents that are characterised as 
equivalence classes by  
\be \hbox{\srm C} \in \langle \hbox{\srm U}\rangle \ \  \Leftrightarrow \ \
u_{_{\rm C}}^{\,\mu}=u_{_{\langle U\rangle}}^{\,\mu}\, .\label{72}\fe
In particular we shall use the label $\langle\emptyset\rangle$
for the class of constituents that are convected with the entropy,
so that 
\be \hbox{\srm C} \in \langle\emptyset\rangle\ \ \Leftrightarrow \ \
v_{_{\rm C}\emptyset}^\mu=0\, .\label{73}\fe

For each such class of comoving constituents, it will be useful
to define combined values of additive quantities such as force density
and stress using notation illustrated by the examples
\be \tilde f_{\,\mu}^{_{\langle U\rangle}}=  
{_\sum\atop ^{_{{\rm C}\in \langle U\rangle}}} 
\tilde f_{\,\mu}^{_{\rm C}}\, ,\hskip 1 cm 
 f_{\!_{\rm ext}\,\mu}^{_{\langle U\rangle}}=  
{_\sum\atop ^{_{{\rm C}\in \langle U\rangle}}} 
f_{\!_{\rm ext}\,\mu}^{_{\rm C}}\, ,\hskip 1 cm
\tau_{\ \ \nu}^{_{\langle U\rangle}\,\mu}=  
{_\sum\atop ^{_{{\rm C}\in \langle U\rangle}}} 
\tau_{\,\nu}^{_{\rm C}\,\mu}\, ,\label{74}\fe
 and more particularly for the ordinary dynamical 4-force density by
\be f^{_{\langle U\rangle}}_{\,\mu}=
{_\sum\atop ^{_{{\rm C}\in \langle U\rangle}}}
f^{_{\rm C}}_{\,\mu}= 2 u_{_{\langle U\rangle}}^{\,\nu}
{_\sum\atop ^{_{{\rm C}\in \langle U\rangle}}}
n_{_{\rm C}}\nabla_{\![\nu}\pi^{_{\rm C}}{_{\!\mu]}}+
{_\sum\atop ^{_{{\rm C}\in \langle U\rangle}}}
\pi{^{_{\rm C}}_{\,\mu}}\nabla_{\!\nu}
n_{_{\rm C}}^{\,\nu}\, .\label{75}\fe

Using this notation we can regroup the terms on the right of
the general entropy creation formula (\ref{42}) in the form
\be {\cal Q} +\Theta \nabla_{\!\mu} s^\mu=-{_\sum \atop ^{_{\rm A}}}
{\cal E}^{_A}_{\!\{\emptyset\}} \nabla_{\!\nu}n_{_{\rm A}}^{\,\nu}
-{_\sum \atop ^{_{\langle U\rangle}}}
\tau_{\ \ \nu}^{_{\langle U\rangle}\,\mu}\nabla_{\!\mu}
u_{_{\langle U\rangle}}^{\,\nu}
-{_\sum \atop ^{_{\langle U\rangle}}} 
v_{_{\!\langle U\rangle}\emptyset}^{\ \,\mu} 
\,\tilde{f} {^{_{\langle U\rangle}}_{\!\{0\}\,\mu}}
-{_\sum \atop ^{^{\rm I}}} 
v_{_{\rm I}\emptyset}^{\ \mu} 
\,{f} {^{_{\rm I}}_{\!\{0\}\,\mu}}
 \, .\label{120}\fe
using the notation introduced above, according to which
\be \,\tilde{f} {^{_{\langle U \rangle}}_{\!\{0\}\mu}}=
f {^{_{\langle U\rangle}}_{\!\{0\}\mu}}+\nabla_{\!\nu}
\tau_{\,\mu}^{_{\langle U\rangle}}{^\nu}- 
f^{_{\langle U\rangle}}_{\!_{\rm ext}\,\mu}\, ,\label{120c}\fe

The first term on the right of (\ref{120}) is the rate of chemical energy 
release which we deal with exactly as before in (\ref{36}) by setting
\be \nabla_{\!\nu}n_{_{\rm A}}^{\, \nu}=-{_\sum\atop ^{_{\rm B}}}\Xi_{_{AB}}
{\cal E}^{_B}_{\{\emptyset\}} \, ,\label{121}\fe
where $\Xi_{_{\rm AB}}$ is the same reactivity matrix as was introduced
above.  The second term on the right of (\ref{120}) is the viscous
energy dissipation rate which we deal with in the same manner as 
in (\ref{55}) by setting
 \be\tau^{_{\langle U\rangle}\,\mu\nu}=-2{_\sum\atop ^{_{\langle V\rangle}}}
\eta^{_{\langle U\rangle\langle V\rangle}}
\sigma_{_{\langle V\rangle}}^{\,\mu\nu}
-{_\sum\atop ^{_{\langle V\rangle}}}
\zeta^{_{\langle U\rangle\langle V\rangle}}\theta_{_{\langle V\rangle}}
\gamma^{\,\mu\nu}\, ,\label{122}\fe
the only difference being that the positive shear viscosity coefficients
$\eta^{_{\langle U\rangle\langle V\rangle}}=
\eta^{_{\langle V\rangle\langle U\rangle}}$ and bulk velocity coefficients 
$\zeta^{_{\langle U\rangle\langle V\rangle}}=
\zeta^{_{\langle V\rangle\langle U\rangle}}$ now only need to be specified 
for the restricted range of the index $\langle${\srm U}$\rangle$ labelling the 
comoving equivalence classes rather than for the full range of the normal 
constituent label {\srm C}. It will similarly be sufficient to specify
just a restricted range of positive resistivity coefficients
$Z^{_{\langle U\rangle\langle V\rangle}}=
Z^{_{\langle V\rangle\langle U\rangle}}$ for 
$\langle${\srm U}$\rangle\neq\langle${\srm V}$\rangle$ to specify the 
resistivity contribution in a force formula whereby the terms in the original
expression (\ref{60}) are regrouped in the form
\be \gamma^{\mu\nu}\,\tilde f{^{_{\langle U\rangle}}_{\!\{0\}\,\nu}}
={_\sum\atop ^{_{\langle V\rangle }}}
f^{_{\langle U\rangle\langle V\rangle}\mu}
+{_\sum\atop ^{_{\rm I}}} f^{_{\langle U\rangle{\rm I}}\,\mu}
\, ,\label{123}\fe
in which the resistive force density terms are given by
\be f^{_{\langle U\rangle\langle V\rangle}\,\mu}=
Z^{_{\langle U\rangle\langle V\rangle}} 
v_{\!_{\langle V\rangle\langle U\rangle}}^{\ \ \mu} \, .\label{124}\fe

It remains to specify the vortex drag terms 
$ f^{_{{\rm I}\langle U\rangle}\,\mu} = -
f^{_{_{\langle U\rangle{\rm I}}}\,\mu}$, which combine to give the
space convected superfluid force densities as
\be \gamma^{\mu\nu} \,{f} {^{_{\rm I}}_{\!\{0\}\,\nu}}
={_\sum\atop ^{_{\langle U\rangle}}} f^{_{{\rm I}\langle U\rangle}\,\mu}
\, ,\label{125}\fe
so that the last two terms in (\ref{120}) can be recombined in the form
\be -{_\sum \atop ^{_{\langle U\rangle}}} 
v_{\!_{\langle U\rangle}\emptyset}^{\ \mu} 
\,\tilde{f} {^{_{\langle U\rangle}}_{\!\{0\}\,\mu}}
-{_\sum \atop ^{_{\rm I}}} v_{_{\rm I}\emptyset}^{\,\mu} 
\,{f} {^{_{\rm I}}_{\!\{0\}\,\mu}}
=\frac{_1}{^2}{_\sum\atop ^{_{\langle U\rangle,\langle V\rangle}}}
v_{_{\langle V\rangle\langle U\rangle}\,\mu} 
 f^{_{\langle U\rangle\langle V\rangle}\,\mu}+
{_\sum\atop ^{_{\langle U\rangle,{\rm I}}}}
v_{\!_{\langle U\rangle{\rm I}}\,\mu}
f^{_{{\rm I}\langle U\rangle}\,\mu} \, .\label{125a}\fe

According to the reasonning of the preceeding section, these contributions 
should be given by expressions of the form
 \be f^{_{{\rm I}\langle U\rangle}\,\mu}
=n_{_{\rm I}}\gamma^{\mu\nu}
\varpi^{_{\rm I}}_{\,\nu\rho}
V_{_{\rm I}}^{_{\langle U\rangle}\,\rho}\, ,\label{107b}\fe
for a set of generalised velocity vectors that add up to give a sum
\be V_{_{\rm I}}^{\,\mu}=
{_\sum\atop ^{_{\langle V\rangle }}}V_{_{\rm I}}^{_{\langle U\rangle}\,\mu}
\, ,\label{107c}\fe
in terms of which we shall get
\be \varpi^{_{\rm I}}_{\,\mu\nu}\big(u_{_{\rm I}}^{\,\mu}+
V_{_{\rm I}}^{\,\mu}\big)=0\, \, ,\label{125c}\fe
and 
\be \,{f} {^{_{\rm I}}_{\!\{0\}\,\mu}}=\varpi^{_{\rm I}}_{\,\mu\nu}
V_{_{\rm I}}^{\,\mu}\, .\label{125d}\fe
The final drag dissipative term in (\ref{125a}) can thereby be rewritten as
\be {_\sum\atop ^{_{\langle U\rangle,{\rm I}}}}
v_{\!_{\langle U\rangle{\rm I}}\,\mu} f^{_{{\rm I}\langle U\rangle}\,\mu} 
={_\sum\atop ^{_{\langle U\rangle,{\rm I}}}} n_{_{\rm I}}
u_{_{\langle U\rangle}}^{\ \mu}\varpi^{_{\rm I}}_{\,\mu\nu}
V_{_{\rm I}}^{_{\langle U\rangle}\,\nu}\, ,\label{125e}\fe

By the same reasonning as in the preceding section, we can ensure
the required positivity of the total (\ref{125}) by adoption of
an ansatz of the form
\be V_{_{\rm I}}^{_{\langle U\rangle}\,\mu}
=-c_{_{\rm J}}^{_{\,\langle U\rangle}}( w^{_{\rm J}})^{-1}
\gamma^{\mu\nu}\varpi^{_{\rm J}}_{\,\nu\rho}
u_{_{\langle U\rangle}}^{\,\nu}\, ,\label{126a}\fe
which is equivalent to setting
\be f^{_{{\rm J}\langle U\rangle}\,\mu}=c_{_{\rm J}}^{_{\,\langle U\rangle}}
n_{_{\rm J}} w^{_{\rm J}}\perp^{\!_{\rm J}\mu}_{\ \nu}
u_{_{\langle U\rangle}}^{\,\nu}\, ,\label{126}\fe
for a set of positive drag coefficients $c_{_{\rm J}}^{_{\,\langle U\rangle}}$
where {\srm J}, like {\srm I}, ranges over the set of superfluid
index labels.

There is however an extreme limit known as pinning -- the analogue for a 
superfluid constituent of convection in the normal case -- representing 
what will occur if some drag coefficient
$c_{_{\rm H}}^{_{\,\langle U\rangle}}$ say is very large, in which case 
the normal flow vector $u_{_{\langle U\rangle}}^{\ \mu}$ will be 
constrained to lie in the relevant vorticity surface, a requirement that 
is evidently expressible as the condition
\be \varpi^{_{\rm H}}_{\,\mu\nu}
u_{_{\langle U\rangle}}^{\ \nu}= 0\, ,\label{127}\fe
which is evidently sufficient to ensure that the corresponding
dissipation term in the sum (\ref{125e}) will simply vanish.

This can be achieved by taking the term $f^{_{\rm H\langle U\rangle}}{^\mu}$
to be given by the  dissipative drag prescription 
of the form (\ref{126}) that applies to the other contributions 
$f^{_{\rm H\langle V\rangle}}{^\mu}$
for $\langle${\srm V}$\rangle\neq\langle${\srm U}$\rangle$ -- and to all the
corresponding force contributions for the unpinned constituents --  
but instead  by the alternative ansatz
\be V_{_{\rm H}}^{_{\!\langle U\rangle}\, \mu}=
v_{_{\langle U\rangle{\rm H}}}^{\ \mu} 
- {_\sum\atop ^{_{\langle V\rangle\neq \langle U\rangle }}}  
V_{_{\rm H}}^{_{\!\langle V\rangle}\, \mu} \, .\label{133}\fe 
that is equivalent to the formula
\be f^{_{\rm H\langle U\rangle}}{^\mu}=n_{_{\rm H}}\gamma^{\mu\nu}
\varpi^{_{\rm H}}_{\,\nu\rho} v_{_{\langle U\rangle{\rm H}}}^{\ \rho}
-{_\sum\atop ^{_{\langle V\rangle\neq \langle U\rangle} }} 
f^{_{\rm H\langle V\rangle}\,\mu}\, ,\label{130}\fe
which is chosen in such a way as to ensure that, according to (\ref{125}),
the total force acting on the pinned constituent will be given by
\be \gamma^{\mu\nu} \,{f} {^{_{\rm H}}_{\!\{0\}\,\nu}}
=n_{_{\rm H}}\gamma^{\mu\nu}
\varpi^{_{\rm H}}_{\,\nu\rho} v_{_{\langle U\rangle{\rm H}}}^{\ \rho}
\, .\label{131}\fe

For any value of the superfluid constituent index {\srm I}, not 
just for the pinned index value  {\srm H} we have been considering,
knowledge of the contravariant space projection
$\gamma^{\mu\nu}{f} {^{_{\rm I}}_{\!\{0\}\,\nu}}$ will be sufficient
for the complete specification of ${f} {^{_{\rm I}}_{\!\{0\}\,\nu}}$
due to the identity $u_{_{\rm I}}^{\,\mu}
{f} {^{_{\rm I}}_{\!\{0\}\,\mu}}$ which implies that we shall have
\be  {f} {^{_{\rm I}}_{\!\{0\}\,\nu}}=\gamma_{_{\rm I}\,\mu\nu}
\gamma^{\nu\rho}{f} {^{_{\rm I}}_{\!\{0\}\,\rho}}\label{136}\fe
with $\gamma_{_{\rm I}\,\mu\nu}$ defined as usual by
 $\gamma_{_{\rm I}\,\mu\nu} u_{_{\rm I}}^{\,\nu}=0$,
$\gamma_{_{\rm I}\,\mu\nu}\gamma^{\nu\rho}=\delta_\mu^{\,\rho}
-t_\mu u_{_{\rm I}}^{\,\rho}$. The formula (\ref{131}) thus implies
that for the constituent with label {\srm H} that is pinned to the set 
of currents with label  $\langle${\srm U}$\rangle$,
we shall have
\be {f} {^{_{\rm H}}_{\!\{0\}\,\mu}} =n_{_{\rm H}}\big(
\varpi^{_{\rm H}}_{\,\mu\nu} v_{_{\langle U\rangle{\rm H}}}^{\ \nu}
- t_\mu u_{_{\rm H}}^{\,\nu}\varpi^{_{\rm H}}_{\,\nu\rho} 
u_{_{\langle U\rangle}}^{\ \rho}\big)\, ,\label{137}\fe
with the evident implication that  $u_{_{\langle U\rangle}}^{\,\mu}
{f} {^{_{\rm H}}_{\!\{0\}\,\mu}}$ will vanish, and hence by the 
definition (\ref{103b}) that the second term in (\ref{137})
will drop out. Thus we are finally left with the simple formula
\be {f} {^{_{\rm H}}_{\!\{0\}\,\mu}} =n_{_{\rm H}}
\varpi^{_{\rm H}}_{\,\mu\nu} v_{_{\langle U\rangle{\rm H}}}^{\ \nu}
\, ,\label{138}\fe
in which the right hand side is interpretable as a gauge
invariant version of the Joukowski formula for the Magnus effect. 
It is easy to see -- using the definition (\ref{103b}) and the
decomposition $u_{\langle U\rangle}^{\ \nu}=u_{_{\rm H}}^{\ \nu}
+v_{\langle U\rangle_{\rm H}}^{\ \nu}$ --  that the application 
of this force law (\ref{138}) is indeed equivalent to the imposition 
of the pinning condition (\ref{127}).

It is to be remarked that the phenomena of pinning and convection
are physically rather similar in that they both can both b
considered as constraints representing the effect of extremely
strong dissipative coupling. However the way they have been dealt
with here mathematically is very different. In the case of pinning
the dynamical equations have been adjusted in such a way that,
after having been imposed as as an initial value restriction,
the relevant restrain will be preserved by the equations of motion.
On the other hand in the case of convection
the constraint has been directly imposed at an algebraic level,
so as to reduce the number of independent components of the system
from 4N, where N is the number of constituents (each with its
own current 4-vector $n_{_{\rm X}}^{\, \mu}$) to 4N-3N' where
N' is the number of independent comotion constraints (each of which
removes the corresponding 3-velocity components from the list
of independent variables while leaving the corresponding number
density). The most familiar example is that of a generic non
barotropic fluid, as characterised by just a single independent velocity,
so that N'=N-1 and the number of independent components
is just N+3, including as a special case the barotropic
fluid model characterised by N=1, N'=0, for which the number
of independent conmponents redices to 4.

\bigskip
{\bf Acknowledgements}
\medskip

The authors wish to thank Silvano Bonazzola, Eric Gourgoulhon, 
David Langlois, Reinhard Prix and David Sedrakian for instructive 
conversations.

\bigskip
{\bf Appendix. The basic convective superconducting superfluid model}
\medskip

In order to set up a large scale neutron star model for the purpose of
describing the pulsar glitch phenomenon --  which is generally recognised 
to depend on relative motion of a superfluid constituent relative to a 
normal background -- the usual kind of perfect fluid model will evidently 
be inadequate. On the other hand it may be hoped that a satisfactory 
description will be attainable without recourse to the very elaborate 
kind of model involving separate allowance for the many degrees of freedom 
(such as those of the electomagnetic field that plays an essential role in 
the mechanics of the external magnetosphere) that would need to be taken 
into account in a highly accurate treatment. As a reasonable compromise, 
for use in such a context, the following  basic convective superconducting 
superfluid kind of model would seem to be appropriate.

The proposed basic model has three independent constitituents of which 
one is superfluid while the other two are subject to a convection
constraint. Thus in the notation of the previous section it is
characterised by N=3 and N'=1 which means that it has 9 dynamically 
independent components.

In the context for which it is intended, the three constituent currents 
are to be considered as consisting of an entropy current 
\be s^\mu=s u_\emptyset^{\,\mu}\, ,\label{140} \fe
  a superfluid neutron current 
\be n_{\rm n}^{\, \mu}=n_{\rm n} u_{\rm n}^{\, \mu}\, ,\label{141}\fe
and a normal current  
\be n_{\rm c}^{\, \mu}=n_{\rm c} u_{\rm c}^{\, \mu}\, ,\label{142}\fe
that convects the entropy flux, with which it shares the unit flow 
4-vector 
\be u_{\rm c}^{\,\mu}=u_\emptyset^{\, \mu}\, ,\label{143}\fe
and that is to be interpreted as 
representing the flux of all other baryons. This normal baryon current 
is to be thought of as consisting not just of protons but at deeper 
levels also of hyperons, whose charge is neutralised by an ambient lepton
gas consisting not just of electrons but at deeper levels also of muons.
In the crust layers the normal baryon constituent will also include the 
neutrons that are confined within atomic nuclei. A realistic treatment 
of the crust (whose outer layers, at densities below $10^{11}$ g/cm$^3$, 
contain no unconfined superfluid neutrons at all) would need the use 
of a model of a more elaborate elastic conducting solid kind, whose 
formulation, in the covariant Newtonian approach developped here, will 
be left for future work. A Newtonian treatment will in any case be 
inadequate for an accurate treatment of the deeper levels, for which a 
relativistic version~\cite{LSC98} of the model would of course be needed. 

As in the relativistic version, there will be a baryon conservation law 
having the form
\be \nabla_{\!\nu}n_{\rm n}^{\, \nu}+
  \nabla_{\!\nu}n_{\rm c}^{\, \nu}=0\, . \label{144}\fe
In the Newtonian approximation, with which we are concerned here, this law 
is to be interpreted as representing the conservation of rest mass, on 
the understanding that both -- superfluid neutron and normal (protonic 
or other) kinds of baryon are treated as having the same rest mass, 
$m$ say, per particle, while there is of course no rest mass
associated with the entropy, i.e. we have 
\be m^\emptyset =0\, ,\hskip 1 cm m^{\rm c}= m\, ,\hskip  1 cm
 m^{\rm n}=m \, . \label{145}\fe
Within the few per cent level of accuracy -- the most that can be 
expected from a Newtonian treatment in this context -- this common rest 
mass $m$ can be chosen indifferently  to be either the mass of the hydrogen 
atom or simply the bare proton mass $m_{_{\rm p}}$ (not to mention the
value traditionally preferred by chemists, which is one sixteenth of the
mass of an ordinary oxigen atom).. 

In order to characterise a particular model of this type, the
essential element -- which is all that is needed in the conservative
case -- is the  pressure function $\Psi$ or equivalently its
dynamical conjugate the  master function $\Lambda_{_{\rm int}}$,
whose unconstrained version is given as a function of the  4-vectors 
$s^\mu$, $n_{\rm c}^{\, \mu}$ and  $n_{\rm n}^{\, \mu}$ in a gauge 
invariant manner -- meaning that it can depend only on the the 6 scalars 
consisting of the number densities $s$  $n_{\rm c}$ and $n_{\rm n}$
together with the relative velocity magnitudes $v_{{\rm c}\emptyset}$
$v_{{\rm n}\emptyset}$ and $v_{\rm nc}$ from which the 
corresponding internal momentum contributions are obtainable as
the partial derivatives given by the variation law
\be \delta\Lambda_{_{\rm int}}= \Theta_\nu \delta s^\nu+
\chi^{\rm c}_{\,\nu} \delta n_{\rm c}^{\,\nu}+
\chi^{\rm n}_{\,\nu} \delta n_{\rm n}^{\,\nu}\, . \label{146}\fe
The pressure function is obtainable from the master function, and
vice versa, by a Legendre type tranformation expressible as
\be \Psi=\Lambda_{_{\rm int}}- \Theta_\nu s^\nu-
\chi^{\rm c}_{\,\nu } n_{\rm c}^{\,\nu}-
\chi^{\rm n}_{\,\nu} n_{\rm n}^{\,\nu}\, , \label{147}\fe
so that its variation law will have the form
\be \delta\Psi= - s^\nu\delta\Theta_\nu - n_{\rm c}^{\,\nu}
\delta\chi^{\rm c}_{\,\nu}- n_{\rm n}^{\,\nu}
\delta\chi^{\rm n}_{\,\nu}\, . \label{148}\fe

The associated stress momentum energy density tensor will be
given~\cite{CCII} by a formula of the standard form
\be T^\mu_{\ \nu}= s^\mu\,\pi^\emptyset_\nu +n_{\rm c}^{\,\mu}\,
\pi^{\rm c}_{\,\nu}+n_{\rm n}^{\,\mu}\,
\pi^{\rm n}_{\,\nu}+\Psi \delta^\mu_{\,\nu}\, . \label{149}\fe
in terms of the corresponding set of complete 4-momentum covectors, 
of which that of the entropy is simply
\be \pi^\emptyset_{\,\mu}=\Theta_\mu\, . \label{150}\fe
while those of the massive constituents are given in terms of
the corresponding frame dependent 3-velocity covectors
$v_{{\rm c}\,\mu}=\gamma_{\mu\nu}u_{\rm c}^{\,\nu}$ and 
$v_{{\rm n}\,\mu}=\gamma_{\mu\nu}u_{\rm n}^{\,\nu}$ by
\be \pi^{\rm c}_{\,\mu}= \chi^{\rm c}_{\,\mu}+m v_{{\rm c}\,\mu} 
- m\big(\frac{_1}{^2} v_{\rm c}^{\,2}+\phi\big)t_\mu\, , \label{150b}\fe
\be \pi^{\rm n}_{\,\mu}= \chi^{\rm n}_{\,\mu}+m v_{{\rm n}\,\mu} 
- m\big(\frac{_1}{^2} v_{\rm n}^{\,2}+\phi\big)t_\mu\, , \label{150c}\fe
where $\phi$ is the gravitational potential. 

In terms of the corresponding thermal, normal, and superfluid vorticity 
forms, namely
\be \varpi^\emptyset_{\,\mu\nu}=2\nabla_{\![\mu} \pi^\emptyset_{\,\nu]} 
\, ,\hskip 1 cm \varpi^{\rm c}_{\,\mu\nu}= 2\nabla_{\![\mu} 
\pi^{\rm c}_{\,\nu]} \, ,\hskip 1 cm \varpi^{\rm c}_{\,\mu\nu}
=2\nabla_{\![\mu} \pi^{\rm c}_{\,\nu]} \, .\label{152}\fe
the associated 4-force covectors will be expressible by the 
defining formulae
\be f^\emptyset_{\,\mu}=s^{\,\nu}\varpi^\emptyset_{\,\nu\mu}+
\Theta_{\mu}\nabla_{\!\nu}s^{\,\nu}\, ,\label{153}\fe
\be f^{\rm c}_{\,\mu}=n_{\rm c}^{\,\nu}\varpi^{\rm c}_{\,\nu\mu}+
\pi^{\rm c}_{\,\mu}\nabla_{\!\nu}n_{\rm c}^{\,\nu}\, ,\label{153b}\fe
\be f^{\rm n}_{\,\mu}=n_{\rm n}^{\,\nu}\varpi^{\rm n}_{\,\nu\mu}+
\pi^{\rm n}_{\,\mu}\nabla_{\!\nu}n_{\rm n}^{\,\nu}\, .\label{153c}\fe
It is to be remarked that if we wanted to describe the superfluid on a 
mesoscopic scale (large compared with the microscopic particle separation 
lengthscales but small compared with the intervortex spacing) we would 
need to impose the restraint that the superfluid vorticity 
$\varpi^{\rm n}_{\,\nu\mu}$ should vanish, but that it will in general 
have a non zero value on the macroscopic scale (meaning one that is large 
compared with the intervortex spacing) for which the present treatment is 
intended.

If we were dealing with a conducting (as opposed to convective) model
the information needed to characterise the dynamical evolution of
the system would consist of a complete specification of all three
of the 4-force covectors that have just been listed, which is equivalent
to a the specification of the three creation rates 
$\nabla_{\!\nu}s^{\,\nu}$, $\nabla_{\!\nu}n_{\rm c}^{\,\nu}$, 
$\nabla_{\!\nu}n_{\rm n}^{\,\nu}$, and of the three corresponding
space projected 3-force vectors $\gamma^{\mu\nu}f^\emptyset_{\,\nu}$
$\gamma^{\mu\nu}f^{\rm c}_{\,\nu}$, $\gamma^{\mu\nu}f^{\rm n}_{\,\nu}$
(the simplest possibility being that of the strictly conservative
case for which all three 4-force covectors are set to zero).
In a convecting model of the kind we wish to consider here, we still
neeed to specify the creation rates $\nabla_{\!\nu}s^{\,\nu}$, 
$\nabla_{\!\nu}n_{\rm c}^{\,\nu}$, $\nabla_{\!\nu}n_{\rm n}^{\,\nu}$,
as well as the space projected 3-force vector
$\gamma^{\mu\nu}f^{\rm n}_{\,\nu}$ of the superfluid constituent,
but in view of the constraint (\ref{143}) we do not need a separate
specification for the corresponding thermal and normal baryon contributions,
but only for their sum $\gamma^{\mu\nu}f^{\langle {\rm c}\rangle}_{\,\nu}$
where $f^{\langle {\rm c}\rangle}_{\,\nu}$ is the combined
4-force density defined by
\be f^{\langle {\rm c}\rangle}_{\ \nu}=f^{\rm c}_{\,\nu}
+f^\emptyset_{\,\nu}\, .\label{155}\fe

The foregoing variational specification of the separate 4-momenta
as functions of the corresponding currents requires that the
master function be defined not just for convectively constrained
configurations but even when there is a relative motion between
the entropy current and the normal baryon constituent.
Such a general, unconstrained specification will indeed be available
if the convective  (9 component) model under consideration has been obtained
as a high resistivity  approximation from an unconstrained (12 component) 
conducting in which all three constituents move independently.
It will however be more economical from a mathematical point of
view to avoid the introduction of redundant information from
such an unconstrained ancestor model, and to work entirely within
the framework of a reduced variational framework in which the
master function $\Lambda_{_{\rm int}}$ and the associated
pressure function $\Psi$ are specified only for the range of
variables allowed by the convectivity constraint (\ref{143}).

In such a reduced formulation, the 9 independent components can be
taken to consist of the 4 components of the superfluid current
vector $n_{\rm n}^{\,\nu}$, the 4 components of the normal baryon current
vector $n_{\rm c}^{\,\nu}$, together with just the density $s$
of the entropy current, whose velocity is not independent but given
by that of the normal baryon current. The most general variation that
is allowed within this reduced formulation will be given by an 
expression of the form 
\be \delta\Lambda_{_{\rm int}}= -\Theta\,\delta s+
\chi^{\langle{\rm c}\rangle}_{\,\mu}\, \delta n_{\rm c}^{\,\nu}+
\chi^{\rm n}_{\,\mu}\, \delta n_{\rm n}^{\,\nu}\, , \label{156}\fe
which does not provide a specification of the separate thermal
and normal baryon momentum covectors $\Theta_\mu$ and 
$\chi^{\rm}_{\,\mu}$ but only of the scalar temperature $\Theta$
and the amalgamated normal momentum covector 
$\chi^{\langle{\rm c}\rangle}_{\ \mu}$
that can be evaluated within the ancestral unconstrained framework as
\be \Theta=-u_\emptyset^{\,\nu}\Theta_{\,\nu}\, ,\hskip 1 cm
\chi^{\langle{\rm c}\rangle}_{\,\mu}=\chi^{\rm c}+\frac{s}{n_{\rm c}}
\big(\Theta t_\mu +\Theta_\mu\big)\, .\label{157}\fe
The corresponding complete amalgamated normal momentum covector
\be \pi^{\langle{\rm c}\rangle}_{\ \mu}=\pi^{\rm c}_{\,\mu}
+\frac{s}{n_{\rm c}}\big(\Theta t_\mu +\Theta_\mu\big)\, ,\label{158}\fe
will be obtainable  in the reduced formulation as
\be \pi^{\langle{\rm c}\rangle}_{\ \mu}=
\chi^{\langle{\rm c}\rangle}_{\ \mu}+m v_{{\rm c}\,\mu} 
- m\big(\frac{_1}{^2} v_{\rm c}^{\,2}+\phi\big)t_\mu\, .\label{159}\fe

It is to be emphasised that it is possible for physically different
ancestor  models -- as characterised by different values
of $\Theta_\mu$ and $\chi^{\rm}_{\,\mu}$ for given values of the
independent currents -- to engender the same reduced model, 
in the sense of providing the same values 
for $\Theta$ and $\chi^{\langle{\rm c}\rangle}_{\,\mu}$,
and hence also for the pressure function (\ref{148}) and the stress
energy tensor (\ref{159}), which will be expressible as
\be \Psi=\Lambda_{_{\rm int}}+ \Theta s
-\chi^{\langle{\rm c}\rangle}_{\ \nu }\, n_{\rm c}^{\,\nu}-
\chi^{\rm n}_{\,\nu} n_{\rm n}^{\,\nu}\, , \label{160}\fe 
and
\be T^\mu_{\ \nu}= n_{\rm c}^{\,\mu}\,\pi^{\langle{\rm c}\rangle}_{\ \nu}
-\Theta s^\mu t_\nu +n_{\rm n}^{\,\mu}\,\pi^{\rm n}_{\,\nu}
+\Psi \delta^\mu_{\,\nu}\, . \label{161}\fe
The formalism of the reduced formulation is not quite so elegant, but it 
has the avantage of avoiding the introduction of operationally
redundant information singling out some particular one of the
compatible unconstrained ancestor models. In the framework of the 
reduced formulation the combined force (\ref{155}) will be expressible as
\be f^{\langle {\rm c}\rangle}_{\ \mu}=2n_{\rm c}^{\,\nu}\nabla_{\![\nu}
\pi^{\langle{\rm c}\rangle}_{\ \mu]}+
\pi^{\langle{\rm c}\rangle}_{\ \mu}\nabla_{\!\nu}n_{\rm c}^{\,\nu}
+s\nabla_\mu\Theta-t_\mu\nabla_{\!\nu}\big(\Theta s^\nu\big)
\, .\label{162}\fe
We thereby obtain an expression of the form
\be \gamma^{\mu\nu} f^{\langle {\rm c}\rangle}_{\!\{0\} \nu}=
2\gamma^{\mu\rho}n_{\rm c}^{\,\nu}\nabla_{\![\nu}
\pi^{\langle{\rm c}\rangle}{_{\!\rho]}}+s\gamma^{\mu\nu}\nabla_\mu\Theta
+\Big(\gamma^{\mu\nu}\big(\chi^{\langle{\rm c}\rangle}_{\ \nu}
-\chi^{\rm n}_{\ \nu}\big)+m\, v_{\rm cn}^{\,\mu}\big)
\nabla_{\!\nu}n_{\rm c}^{\,\nu}\, ,\label{162b}\fe
for the corresponding adjusted (gauge invariant) 3-force density vector, 
whose analogue, for the free neutron current, will be given simply by
\be \gamma^{\mu\nu} f^{\rm n}_{\!\{0\} \nu}=\gamma^{\mu\rho}
n_{\rm n}^{\,\nu}\varpi^{\rm n}_{\,\nu\rho}\, ,\hskip 1 cm
\varpi^{\rm n}_{\,\nu\rho}=2\nabla_{\![\nu}
\pi^{\rm n}{_{\!\rho]}}\, .\label{162c}\fe

In order to complete the determination of the dynamical equations of
the model, it is necessary to choose the rules specifying the values
of these space projected 3-force vectors $\gamma^{\mu\nu}
f^{\langle {\rm c}\rangle}_{\!\{0\}\nu}$ and 
$\gamma^{\mu\nu} f^{\rm n}_{\!\{0\} \nu}$, and of the two
independent creation rates $\nabla_{\!\nu}s^{\,\nu}$
and $\nabla_{\!\nu}n_{\rm n}^{\,\nu}$ -- of which the latter, by  
(\ref{144}), determines   $\nabla_{\!\nu}n_{\rm c}^{\,\nu}$.
The simplest possibility is of course that of a strictly conservative 
model for which the forces and creation rates all vanish. What
we want to consider here is the more general case in which there
is internal dissipation by the mechanisms described in the
preceeding sections and perhaps also a non vanishing heat loss rate
${\cal Q}$ (due to neutrino emission) but we shall suppose that
although the system may thus be open in the thermodynamic sense
it is nevertheless isolated in the sense there are no external contributions
to the space projected force densities, nor to the time component
of the force acting on the free neutron current, so that in the notation
of the preceding section we shall have
\be f^{\rm n}_{\!_{\rm ext\,} \nu}=0 \, ,\hskip 1 cm
f^{\langle {\rm c}\rangle}_{\!_{\rm ext\,} \nu}
={\cal Q}t_\nu\, .\label{164}\fe
According to the general principles developed in the preceding sections, 
the space projected forces will therefore be given by expressions of the 
standard form  
\be \gamma^{\mu\nu} f^{\langle {\rm c}\rangle}_{\!\{0\} \nu}=
f^{\langle{\rm c}\rangle{\rm n}\mu}-\nabla_{\!\nu}\,
\tau^{\langle{\rm c}\rangle\,\mu\nu}\, ,\label{165a}\fe
\be \gamma^{\mu\nu} f^{\rm n}_{\!\{0\}\nu}=-
f^{\langle{\rm c}\rangle{\rm n}\mu}-\nabla_{\!\nu}\,
\tau^{{\rm n}\,\mu\nu}\, .\label{165b}\fe
In a high temperature version of the model, the mutual
interaction force density $f^{\langle{\rm c}\rangle{\rm n}\mu}$ would be
given in terms of an ordinary positive resistivity coefficient 
$Z^{\langle{\rm c}\rangle{\rm n}}$ by an expression of the form
\be f^{\langle{\rm c}\rangle{\rm n}\mu}=Z^{\langle{\rm c}\rangle{\rm n}}
v_{\rm cn}^{\ \mu}\, .\label{166}\fe
What we are particularly interested in here however  is the low temperature
version of the model, in which the free neutron current is a superfluid,
which means that instead of being given by an ordinary resistivity
formula of the form (\ref{166}) the mutual interaction will be given 
by a vortex drag formula of the kind given by the formula (\ref{126}).

In the present case, the vortex drag force density will be given by
the expression
\be f^{\langle{\rm c}\rangle{\rm n}\,\mu} =-n_{\rm n}\gamma^{\mu\nu}
\varpi^{\rm n}_{\,\nu\rho}V_{\rm n}^{\langle{\rm c}\rangle\,\rho}
\, ,\label{167}\fe
in which $c_{\rm n}^{\,\langle{\rm c}\rangle}$ is the positive
drag coefficient, and the vector $V_{\rm n}^{\langle{\rm c}\rangle\,\mu}$
is  defined by 
\be V_{\rm n}^{\langle{\rm c}\rangle\,\mu}=-\frac{
c_{\rm n}^{_{\,\langle{\rm c}\rangle}}}{w^{\rm n}}\gamma^{\mu\nu}
\varpi^{{\rm n}}_{\,\nu\rho}u_{\rm c}^{\,\rho}\, ,\hskip 1 cm
(w^{\rm n})^2={_1\over ^2}\gamma^{\mu\nu}
\gamma^{\rho\sigma}\varpi^{{\rm n}}_{\,\mu\rho}
\varpi^{{\rm n}}_{\,\nu\sigma}\, ,\label{167a}\fe
or alternatively just by
\be V_{\rm n}^{\langle{\rm c}\rangle\,\mu}=
v_{\langle{\rm c}\rangle n}^{\ \mu}\, ,\label{167b}\fe
in the non-dissipative large $c_{\rm n}^{\,\langle{\rm c}\rangle}$ limit 
case of vortex pinning. In any such (pinned or unpinned) superfluid model, 
the free neutron current will not be subject to any viscosity force, i.e. 
we shall have 
\be \tau^{{\rm n}\,\mu\nu}=0\, ,\label{168a}\fe
but for the combined thermal and normal baryon current contribution
there will in general be a non vanishing viscosity contribution
of the standard form
\be \tau^{\langle{\rm c}\rangle\,\mu\nu}=-2\eta^{\langle{\rm c}\rangle}
\sigma_{\rm c}^{\ \mu\nu}-\zeta^{\langle{\rm c}\rangle}
\theta_{\rm c}\gamma^{\ \mu\nu}\, ,\label{168b}\fe
in which $\eta^{\langle{\rm c}\rangle}$ and  $\zeta^{\langle{\rm c}\rangle}$
are positive shear and bulk viscosities (of which the latter will be
negligibly small in many circumstances) and $\sigma_{\rm c}^{\ \mu\nu}$ and 
$\theta_{\rm c}$ are the trace free and trace parts of the normal 
constituent's expansion tensor as given, according to (\ref{52}) by
\be \theta_{\rm c}^{\,\mu\nu}=\gamma^{\sigma(\mu}\nabla_{\!\sigma}
u_{\rm c}^{\,\nu)}= \sigma_{\rm c}^{\,\mu\nu}
+{_1\over^3}\theta_{\rm c}\gamma^{\mu\nu}\, .\label{168c}\fe

To complete the specification of the model it remains to give
the prescription for the  creation rates. In the present case
the only relevant kind of generalised chemical reaction is one whereby
a free neutron is created by a processe such as ``dripping'' out of a 
confined state within a neutron star crust nucleus or by inverse beta 
decay of a proton in the core, so that the corresponding creation numbers 
will be $N_{\rm n}=1$ and $N_{\rm c}=-1$, and the corresponding chemical 
affinity will be given by the formula
\be {\cal A}_{\{\emptyset\}}=\cal E^{\rm c}_{\{\emptyset\}}
-\cal E^{\rm n}_{\{\emptyset\}}\, ,\label{170}\fe
in terms of the relevant thermal rest frame energies as defined in terms 
of the relevant (ancestral or reduced) 4-momentum covectors by
\be {\cal E}^{\rm c}_{\{\emptyset\}}=-u_\emptyset^{\,\nu}
\pi^{\rm c}_{\ \nu}=-u_\emptyset^{\,\nu}
\pi^{\langle{\rm c}\rangle}_{\ \nu}\, ,\hskip 1 cm 
{\cal E}^{\rm n}_{\{\emptyset\}}
=-u_\emptyset^{\,\nu}\pi^{\rm n}_{\ \nu}\, .\label{171}\fe

The superfluid particle creation rate can thus be seen to be given
in terms of the relevant transfusion coefficient $\kappa$ by an expression 
of the standard form
\be \nabla_{\!\nu}n_{\rm n}^{\,\nu}=\kappa \, {\cal A}_{\{\emptyset\}}
\, ,\label{172}\fe
in which the affinity is given in terms of the relative flow velocity
magnitude $v_{\rm nc}$ by the manifestly frame independent formula
\be  {\cal A}_{\{\emptyset\}}=
\chi^{\rm c}_\natural -\chi^{\rm n}_\emptyset - \frac{_1}{^2}
m v_{\rm nc}^{\, 2}\, ,\label{173}\fe
where
\be\chi^{\rm c}_\natural=-u_{\rm c}^{\,\nu} \chi^{\rm c}_{\,\nu}=
 -u_{\rm c}^{\,\nu}\chi^{\langle{\rm c}\rangle}_{\,\nu}\, ,\hskip
1 cm \chi^{\rm n}_\emptyset=-u_\emptyset^{\,\nu}\chi^{\rm n}_{\,\nu}=
-u_{\rm c}^{\,\nu}\chi^{\rm n}_{\,\nu}\, .\label{173a}\fe

The last thing we need to complete the specification of the model
is the value of the energy emission rate ${\cal Q}$ that determines the
entropy rate via the formula (\ref{120}), which gives
\be {\cal Q} +\Theta \nabla_{\!\mu} s^\mu=\kappa {\cal A}_{\{\emptyset\}}^2
+2 \eta^{\langle {\rm c}\rangle}\gamma_{\mu\nu}\gamma_{\rho\sigma}\
\sigma_{\rm c}^{\,\mu\rho}\sigma_{\rm c}^{\,\nu\sigma}+ 
\zeta^{\langle {\rm c}\rangle}\theta_{\rm c}^2+
{n_{\rm n}w^{\rm n}\over c_{\rm n}^{\,\langle {\rm c}\rangle}}
V_{\rm n}^{\langle{\rm c}\rangle\,\mu}\gamma_{\mu\nu}
V_{\rm n}^{\langle{\rm c}\rangle\,\nu}\, .\label{174}\fe
The complete set of dynamical equations for the 9 independent
components (those of the space vectors $v_{\rm c}^{\,\mu}$ and 
$v_{\rm n}^{\,\mu}$ together with the scalars $n_{\rm c}$, $n_{\rm n}$
and $s$) is thus completed: it consists of the creation formulae
(\ref{144}), (\ref{172}) and (\ref{174}), together with
the pair of 3-force equations (\ref{165a}), (\ref{165b}) (as made explicit
by the prescriptions (\ref{167}), (\ref{167a}, (\ref{168a}), and 
 (\ref{168b}) for drag and viscosity).

It is to be remarked that final term in (\ref{174}) will drop out
in the large $c_{\rm n}^{\,\langle{\rm c}\rangle}$ limit for which
the drag prescription (\ref{167a}) is replaced by the vortex pinning
prescription (\ref{167a}). In a similar way the first term in (\ref{174})
in the large $\kappa$ limit for which the system will be maintained
in a state of chemical equilibrium as characterised by the condition
${\cal A}_{\{\emptyset\}}=0$ which will have the effect of reducing the
number of dynamically independent components from 9 to 8.

The particular case of a thermodynamically closed model -- which may
be a good approximation for processes occurring on a short timescale
-- will be obtained by setting ${\cal Q}=0$. However it will often be 
more realistic to take ${\cal Q}$ to have a value that is positive
and monotonically increasing as a function of the temperature
$\Theta$ to allow for losses by URCA type neutrino emission
processes. For processes ocurring over a sufficiently long
timescale the temperature sensitivity of ${\cal Q}$ near some
emission threshold value may be sufficient
to justify the use of a simplifying approximation whereby the
temperature $\Theta$ is held fixed at the threshold value in
question, thereby determining the value of $s$ and hence of the creation
rate $\nabla_{\!\mu}s^\mu$. This will reduce the number of dynamically 
independent components from 9 to 8 (or, in the chemical equilibrium case,
from 8 to 7) so that (\ref{174}) will no longer
be needed as a dynamical equation of the system, but will merely serve
for the purpose of calculating the corresponding value of ${\cal Q}$
in case it might be needed. A simple extreme special case of such a
fixed temperature thermodynamically open variant of the model is the 
zero temperature limit for which $\Theta$ and $s$ both vanish.


\begin{thebibliography}{99}

\bibitem{CCI} B. Carter, N. Chamel, 
``Covariant analysis of Newtonian multi-fluid models for neutron stars: 
I Milne-Cartan structure and variational formulation,''
{\it Int. J. Mod. Phys.} {\bf D13} (2004) 291-326. [astro-ph/0305186]

\bibitem{CCII} B. Carter, N. Chamel, 
``Covariant analysis of Newtonian multi-fluid models for neutron stars: 
II Stress - energy tensors and virial theorems.'' [astro-ph/0312414] 

\bibitem{CK94} B. Carter, I.M. Khalatnikov,
``Canonically covariant formulation of  Landau's Newtonian superfluid 
dynamics'', 
{\it Rev. Math. Phys.} {\bf 6} (1994)  277-304.

\bibitem{Prix04} R. Prix, ``Variational description of multi-fluid 
hydrodynamics: uncharged fluids'', 
{\it Phys.Rev.} {\bf  D69} (2004) 043001. [physics/0209024]

\bibitem{PCA} R. Prix, G.L. Comer, N. Andersson,
``Inertial modes of non-stratified superfluid neutron stars''
{\it Mon. Not. R. Astr. Soc.} {\bf 348} (2004) 652-662. 
[astro-ph/0308507] 

\bibitem{YE04} S. Yoshida, Y. Eriguchi ``Rapidly rotating superfluid 
neutron stars in Newtonian dynamics'', 
{\it Mon. Not. R. Astr. Soc.} {\bf 347} (2004) 575. [astro-ph/0309522]

\bibitem{LM00} L. Lindblom, G. Mendell, `` R-modes in superfluid neutron 
stars'', {\it Phys. Rev.} {\bf D61} (2000) 104003. [gr-qc/9909084]

\bibitem{CLS} B. Carter, D. Langlois, D.M. Sedrakian,
``Centrifugal buoyancy as a mechanism for neutron star glitches'',
{\it Astron. and Astrophys.} {\bf 361} (2000) 795 - 802.
[astro-ph/0004121]
 
\bibitem{C89} B. Carter,
``Covariant Theory of Conductivity in Ideal Fluid or Solid Media",
in {\it Relativistic Fluid Dynamics (C.I.M.E., Noto, May 1987)}
ed.  A.M. Anile, \& Y. Choquet-Bruhat, Lecture Notes in Mathematics {\bf 1385}
(Springer - Verlag, Heidelberg, 1989) 1-64.

\bibitem{C91} B. Carter,
``Convective variational approach to relativistic thermodynamics of 
dissipative fluids''
{\it Proc. Roy. Soc. Lond.} {\bf A433} (1991) 45-62.

\bibitem{LSC98} D. Langlois, D. Sedrakian, B. Carter,
``Differential rotation of relativistic superfluid in neutron stars''
{\it Mon, Not. R. Astr. Soc.} {\bf 297} (1998)  1198-1201.
[astro-ph/9711042] 

\bibitem{IsraelStewart79} W. Israel, J.M. Stewart
{\it Proc. Roy. Soc. Lond.} {\bf A365} (1979) 43.

\bibitem{Prigogine67} I. Prigogine
{Thermodynamcs off irreversible processes}
(Wiley, New York, 1967).


\end{thebibliography}
\end{document}